\documentclass[a4paper,12pt]{article}
\pdfoutput=1
\usepackage{jheppub}

\usepackage{graphicx}
\usepackage{hyperref}
\usepackage{cancel}
\usepackage{amssymb}
\usepackage{textcomp}
\usepackage{amsmath}
\usepackage{bm}
\usepackage{times}
\usepackage{epsfig}
\usepackage{color}

\begin{document}
\title{ Heavy Neutral Leptons in Gauged $U(1)_{L_\mu-L_\tau}$ at Muon Collider}
\bigskip
\author{Ru-Yi He$^1$}
\author{Jia-Qi Huang$^1$}
\author{Jin-Yuan Xu$^1$}
\author{Fa-Xin Yang$^2$}
\author{Zhi-Long Han$^1$}
\emailAdd{sps\_hanzl@ujn.edu.cn}
\author{Feng-Lan Shao$^2$}
\emailAdd{shaofl@mail.sdu.edu.cn}
\affiliation{$^1$School of Physics and Technology, University of Jinan, Jinan, Shandong 250022, China}
\affiliation{$^2$School of Physics and Physical Engineering, Qufu Normal University, Qufu, Shandong 273165, China}

\abstract{ Heavy neutral leptons are the most appealing candidates to generate the tiny neutrino masses. In this paper, we study the signature of heavy neutral leptons in gauged $U(1)_{L_\mu-L_\tau}$ at a muon collider. Charged under the $U(1)_{L_\mu-L_\tau}$ symmetry, the heavy neutral leptons can be pair produced via the new gauge boson $Z'$ at muon collider as $\mu^+\mu^-\to Z^{\prime *}\to NN$ and $\mu^+\mu^-\to Z^{\prime (*)} \gamma\to NN\gamma$. We then perform a detailed analysis on the lepton number violation signature $\mu^+\mu^-\to NN\to \mu^\pm\mu^\pm W^\mp W^\mp$ and $\mu^+\mu^-\to NN \gamma\to \mu^\pm\mu^\pm W^\mp W^\mp \gamma$ at the 3 TeV muon collider, where the hadronic decays of $W$ boson are treated as fat-jets $J$. These lepton number violation signatures have quite clean backgrounds at the muon collider. Our simulation shows that a wide range of viable parameter space is within the reach of the 3 TeV muon collider. For instance, with new gauge coupling $g'=0.6$ and an integrated luminosity of 1000 fb$^{-1}$, the $\mu^\pm\mu^\pm JJ$ signal could probe $m_{Z'}\lesssim 12.5$ TeV. Meanwhile, if the gauge boson mass satisfies $2 m_N<m_{Z'}<\sqrt{s}$, the $\mu^\pm\mu^\pm JJ\gamma$ signature would be more promising than the $\mu^\pm\mu^\pm JJ$ signature.
}

\keywords{Heavy Neutral Lepton, Muon Collider, Lepton Number Violation Signature}

\maketitle

\section{Introduction}

Heavy neutral leptons $N$ are well motivated to explain the origin of tiny neutrino masses. Due to the singlet nature of heavy neutral leptons under the standard model gauge group, we can write a Majorana mass term $m_N \overline{N^c}N$, which results in light Majorana neutrino masses as $m_\nu\sim m_D^2/m_N$ via the type-I seesaw mechanism \cite{Minkowski:1977sc,Mohapatra:1979ia}. To generate the sub-eV neutrino masses, the heavy neutral lepton masses should be at very high scale $m_N\gtrsim10^{14}$~GeV if $m_D$ is at the electroweak scale, which is far beyond current experimental reach. Alternatively, an electroweak scale $m_N$ is usually assumed for phenomenological studies \cite{Abdullahi:2022jlv}, which is then determined by the mixing parameter $V_{\ell N}$.

To confirm the Majorana nature of neutrinos, lepton number violation signatures are expected. The most sensitive experiment is the neutrinoless double beta decay, which could probe the inverted mass ordering scenario in the next generation experiments \cite{Agostini:2022zub}. On the other hand, the detection of the lepton number violation signature at colliders could unravel the explicit mechanism of tiny neutrino masses \cite{Cai:2017mow}. For heavy neutral leptons in the type-I seesaw, the distinct signature at the hadron collider is $pp\to W^{\pm*}\to \ell^\pm N\to\ell^\pm\ell^\pm jj$ \cite{Han:2006ip}. However, the sensitive region of this signature heavily depends on a relatively large mixing parameter $V_{\ell N}$ \cite{CMS:2018jxx,ATLAS:2019kpx}, which is usually much higher than the natural seesaw prediction $V_{\ell N}\sim\sqrt{m_\nu/m_N}$.

Apart from the canonical seesaw there are many theories with extended gauge groups, such as the $U(1)_{B-L}$ model \cite{Mohapatra:1980qe} and the left-right symmetric model \cite{Keung:1983uu}. In these models, the heavy neutral leptons are charged under the extended gauge group, which opens new production mechanisms for the heavy neutral leptons. Search for pair production of heavy neutral leptons via the decay of $Z'$ boson in the framework of the left-right symmetric model has been performed recently by the CMS collaboration \cite{CMS:2023ooo}. The region within $m_N\lesssim1.4$~TeV and $m_{Z'}\lesssim4.4$ TeV has been excluded by the dimuon channel. Meanwhile, the ATLAS collaboration searches for right-handed $W'$ boson decaying to heavy neural lepton $N$ and leptons $\ell$, which can exclude the region with $m_N\lesssim3.8$ TeV and $m_{W'}\lesssim6.4$~TeV \cite{ATLAS:2023cjo}.

The construction of a multi-TeV muon collider is proposed recently \cite{Delahaye:2019omf,Long:2020wfp}.
Since then, searches for heavy neutral leptons at the multi-TeV muon collider have drawn  increasing interest \cite{Liu:2021akf,Li:2022kkc,Chakraborty:2022pcc,Mikulenko:2023ezx,Wang:2023zhh}. One promising signature is $\mu^+\mu^-\to N\nu$, which could probe the mixing parameter $|V_{\mu N}|^2\gtrsim10^{-6}$ at the 10 TeV muon collider \cite{Mekala:2023diu,Kwok:2023dck,Li:2023tbx}. Due to missing neutrino in the final states, this signal can hardly confirm the Majorana nature of heavy neutral leptons. Another interesting signature is the lepton number violation signal via vector boson scattering process $W^\pm Z/\gamma\to \ell^\pm N$ \cite{Li:2023lkl}, via associated production process $\mu^+\mu^-\to N W^\pm \ell^\mp$ \cite{Antonov:2023otp}, or at the same-sign muon collider via $\mu^+\mu^+\to W^+W^+$ process \cite{Jiang:2023mte}. These studied signatures also require that the mixing parameter $V_{\ell N}$ is not too small. Meanwhile, under the tight constraints from current experimental searches \cite{CMS:2023ooo,ATLAS:2023cjo}, the first stage of a 3 TeV muon collider \cite{MuonCollider:2022xlm} is  unpromising to probe heavy neutral leptons in the $U(1)_{B-L}$ and left-right symmetric models.

The multi-TeV muon collider is a perfect machine to test the muon-philic forces. One attractive option is the  gauged $U(1)_{L_\mu-L_\tau}$ model \cite{He:1990pn,Das:2022mmh,Sun:2023rsb}. Compared to the $U(1)_{B-L}$ or left-right symmetry, the $U(1)_{L_\mu-L_\tau}$ symmetry is less constrained due to the lack of direct couplings to electron and quarks. For instance, there is only a loose constraint from neutrino trident production when $m_{Z'}\gtrsim100$ GeV\cite{CCFR:1991lpl}. Therefore, this model is  extensively studied to explain the anomaly of the muon magnetic moment \cite{Muong-2:2021ojo,Baek:2001kca}, $B$ meson anomaly \cite{Altmannshofer:2014cfa,LHCb:2021trn}, dark matter \cite{Baek:2008nz,Holst:2021lzm}, and neutrino masses \cite{Baek:2015mna,Asai:2017ryy}. It is shown that  a 3 TeV muon collider is powerful enough to probe quite a large parameter space with $m_{Z'}\lesssim10$~TeV \cite{Huang:2021nkl,Dasgupta:2023zrh}.
In this paper,  we consider the gauged $U(1)_{L_\mu-L_\tau}$ model with three heavy neutral leptons $N_e,N_\mu,N_\tau$. We then study the pair production of heavy neutral lepton $N$ via $Z'$ boson at a 3 TeV muon collider. To test the Majorana nature of neutrinos, we focus on the lepton number violation process $\mu^+\mu^-\to Z^{\prime *}\to NN\to \mu^\pm\mu^\pm W^\mp W^\mp$ and $\mu^+\mu^-\to Z^{\prime} \gamma\to NN \gamma\to \mu^\pm\mu^\pm W ^\mp W^\mp \gamma$ with the hadronic decay of $W$.

The paper is organized as follows. In Section \ref{SEC:MD}, we review the gauged $U(1)_{L_\mu-L_\tau}$ model with three heavy neutral leptons and discuss relevant experimental constraints. The decay properties of gauge boson $Z'$ and heavy neutral lepton $N$ are also considered in this section. In Section \ref{SEC:PR}, we study the pair production of heavy neutral lepton at the 3~TeV muon collider. Analysis of the lepton number violation signatures are performed in Section~\ref{SEC:LNV}. The conclusion is in Section \ref{SEC:CL}.

\section{The Model}\label{SEC:MD}

In this paper, we consider the anomaly-free gauged $U(1)_{L_\mu-L_\tau}$ extension of the type-I seesaw mechanism. Three heavy neutral leptons $N_e,N_\mu,N_\tau$ with $U(1)_{L_\mu-L_\tau}$ charge $(0,1,-1)$ are introduced to generate the tiny neutrino masses. Predicting type {\bf C}$^R$ of the two-zero minor \cite{Lavoura:2004tu,Araki:2012ip}, the minimal model with only one scalar singlet $\Phi_1$, which carries $U(1)_{L_\mu-L_\tau}$ charge $+1$, is now tightly constrained by the neutrino oscillation data and the sum of light neutrino masses \cite{Asai:2018ocx,Asai:2020qax}. So the second scalar singlet $\Phi_2$ with $U(1)_{L_\mu-L_\tau}$ charge $+2$ is also employed to reduce the above conflict \cite{Borah:2021mri}. The Yukawa interactions and mass terms relevant to neutrino masses are given by \cite{Patra:2016shz}
\begin{eqnarray}
	\mathcal{L}&\supset& - y_e \bar{L}_e \tilde{H} N_e - y_\mu \bar{L}_\mu \tilde{H} N_\mu - y_\tau \bar{L}_\tau \tilde{H} N_\tau -\frac{1}{2} M_{ee} \overline{N^c_e}N_e-M_{\mu\tau}{N^c_\mu}N_\tau \\ \nonumber
	&&-y_{e\mu} \Phi_1^\dagger \overline{N^c_e}N_\mu -y_{e\tau} \Phi_1 \overline{N^c_e}N_\tau-\frac{1}{2}y_{\mu\mu} \Phi_2^\dagger \overline{N^c_\mu}N_\mu -\frac{1}{2}y_{\tau\tau} \Phi_2 \overline{N^c_\tau}N_\tau+ \text{h.c.},
\end{eqnarray}
where $L_e,L_\mu,L_\tau$ are the lepton doublets, $H$ is the standard model Higgs doublet, and $\tilde{H}=i\tau_2 H^*$. After the spontaneous symmetry breaking, we can denote the vacuum expectation values of scalars as $\langle H \rangle=v_0,\langle \Phi_1 \rangle=v_1,\langle \Phi_2 \rangle=v_2$. Then the Dirac neutrino mass matrix and heavy neutral lepton mass matrix are given by
\begin{align}
	M_D=\left(
	\begin{array}{c c c}
		y_e v_0 & 0 & 0\\
		0 & y_\mu v_0&0 \\
		0 & 0 & y_\tau v_0
	\end{array}
	\right), \quad
	M_N =
	\left(
	\begin{array}{c c c}
		M_{ee} & y_{e\mu}v_1 & y_{e\tau}v_1\\
		y_{e\mu}v_1 & y_{\mu\mu} v_2& M_{\mu\tau} \\
		y_{e\tau}v_1 & M_{\mu\tau} & y_{\tau\tau} v_2
	\end{array}\right).
\end{align}
Light neutrino masses are generated via the type-I seesaw formula
\begin{equation}\label{Eq:SS}
	M_\nu\simeq - M_D M_N^{-1} M_D^T.
\end{equation}
Without any specific structure of heavy neutral lepton mass matrix $M_N$, the obtained light neutrino mass matrix $M_\nu$ is a general symmetric matrix, thus can easily fit the neutrino oscillation data \cite{deSalas:2020pgw}. Inversely, we can use Eqn.~\eqref{Eq:SS} to express the mass matrix $M_N$ as \cite{Granelli:2023egb}
\begin{equation}
	M_N=-M_D^T M_\nu^{-1} M_D,
\end{equation}
where $M_\nu=U^*\text{diag}(m_{\nu_1},m_{\nu_2},m_{\nu_3}) U^\dag$, and $U$ is the light neutrino mixing matrix. In this way, the mass matrix $M_N$ is determined by the light neutrino oscillation data and the Yukawa coupling $y_e,y_\mu,y_\tau$. The mass matrix $M_N$ can be diagnalized by a unitarity matrix $\Omega$ as
\begin{equation}
	\Omega^T M_N \Omega = \text{diag}(m_{N_1},m_{N_2},m_{N_3}).
\end{equation}

Considering the mixing with light neutrinos,  the heavy neutral leptons interact with $W^\pm,Z$ gauge boson and Higgs boson as
\begin{equation}\label{Eq:HNL}
	\mathcal{L}\supset -\frac{g}{\sqrt{2}}  \bar{N}_k V_{\ell k}^{*} \gamma^\mu P_L \ell W_\mu 
	-\frac{g}{2\cos\theta_W} \bar{N}_k V_{\ell k}^{*}  \gamma^\mu P_L \nu_\ell Z_\mu   - \frac{g m_N}{2 m_W} \bar{N}_k V_{\ell k}^{*} P_L \nu_\ell h    + \text{h.c.},
\end{equation}
where the mixing matrix $V\simeq M_D M_N^{-1}$. For electroweak scale heavy neutral lepton, we assume a seesaw induced mixing parameter $V_{\ell N}\sim \sqrt{m_\nu/m_N}\sim10^{-6}$, which is allowed by current experimental limits \cite{Abdullahi:2022jlv}.

The interactions of the new boson $Z'$ with fermions are
\begin{equation}
	\mathcal{L}\supset g'(\bar{\mu}\gamma^\mu\mu-\bar{\tau}\gamma^\mu\tau+\bar{\nu}_\mu\gamma^\mu P_L\nu_\mu-\bar{\nu}_\tau\gamma^\mu P_L\nu_\tau+\bar{N}_\mu\gamma^\mu P_R N_\mu-\bar{N}_\tau\gamma^\mu P_R N_\tau)Z'_\mu.
\end{equation}
In terms of mass eigenstates, the new neutral current of neutral leptons can be rewritten as
\begin{equation}\label{Eq:Zp}
	\mathcal{L}\supset g'\left[\bar{\nu}_i (U_{\mu i}^{*}U_{\mu j}-U_{\tau i}^{*}U_{\tau j})\gamma^\mu  P_L\nu_j+\bar{N}_i (\Omega_{\mu i}^{*}\Omega_{\mu j}-\Omega_{\tau i}^{*}\Omega_{\tau j})\gamma^\mu  P_R N_j\right]Z'_\mu,
\end{equation}

From Eqn.~\eqref{Eq:HNL} and Eqn.~\eqref{Eq:Zp}, it is obvious that the collider phenomenology of heavy neutral leptons depends on the mixing patterns, which makes the results model dependent. In the following collider simulation, we consider a muon-philic heavy neutral lepton $N$, which couples exclusively to muon \cite{Drewes:2022akb}. We also take the mixing parameter $\Omega_{\mu N}^{*}\Omega_{\mu N}-\Omega_{\tau N}^{*}\Omega_{\tau N}=1$ in the following calculation, so the general model dependent results can be obtained by a simple rescaling.

\subsection{Constraints}

\begin{figure}
	\begin{center}
		\includegraphics[width=0.5\linewidth]{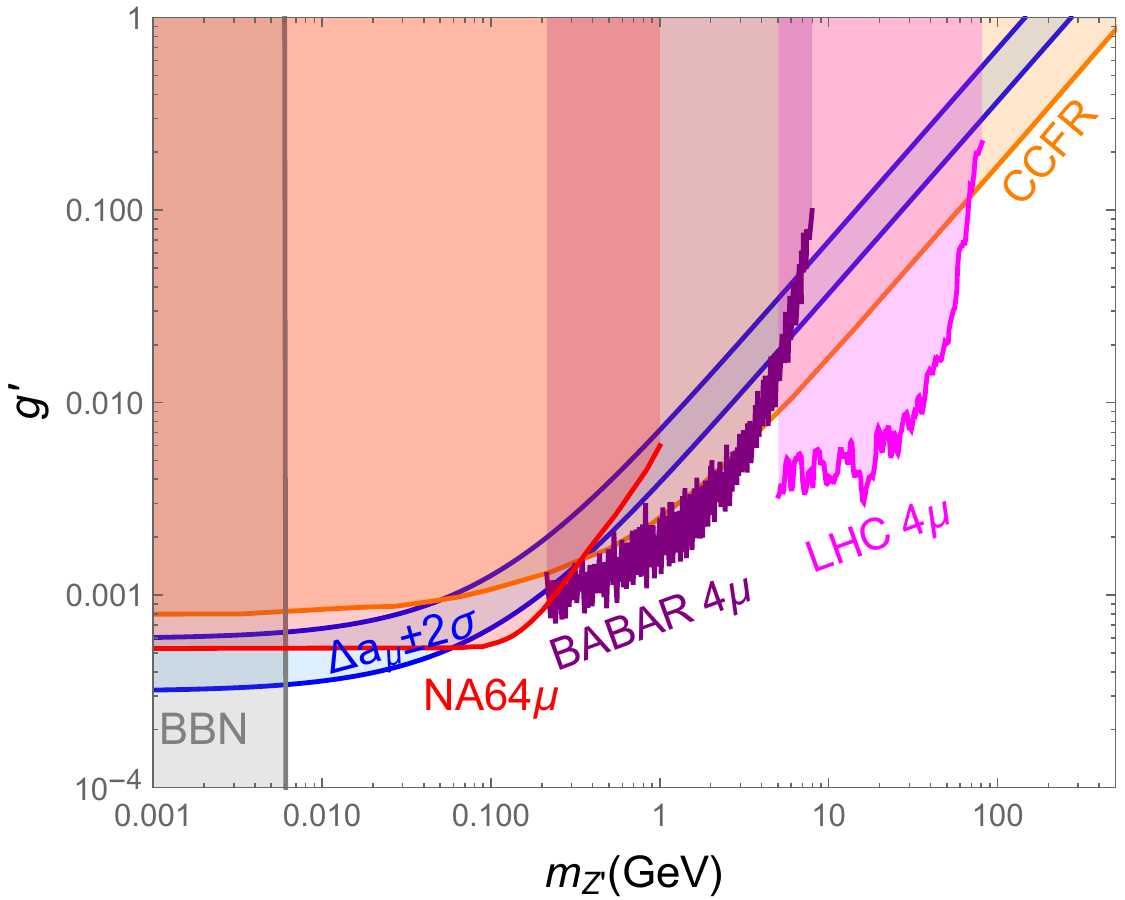}
	\end{center}
	\caption{Constraints on the gauged $U(1)_{L_\mu-L_\tau}$. The blue band can explain the $\Delta a_\mu$ result \cite{Muong-2:2021ojo}. The gray, orange, red, purple, and pink regions are excluded by BBN \cite{Escudero:2019gzq}, CCFR \cite{CCFR:1991lpl}, NA64$\mu$ \cite{Andreev:2024sgn}, BABAR \cite{BaBar:2016sci}, and LHC \cite{ATLAS:2023vxg} respectively.}
	\label{FIG:CS}
\end{figure}

In this section, we briefly summarize the constraints on the gauged $U(1)_{L_\mu-L_\tau}$ model. One important motivation of this symmetry is to explain the anomaly of the muon magnetic moment $\Delta a_\mu=(251\pm59)\times10^{-11}$ \cite{Muong-2:2021ojo}. The new gauge boson $Z'$ contributes to $\Delta a_\mu$ at one loop level, which is evaluated as \cite{Baek:2001kca}
\begin{equation}
	\Delta a_\mu=\frac{g^{\prime 2}}{4\pi^2}\int_0^1dx\frac{m_\mu^2 x(1-x)^2}{m_\mu^2(1-x)^2+m_{Z'}^2 x}.
\end{equation}
The viable parameter space to interpret $\Delta a_\mu$ is shown in Figure~\ref{FIG:CS}. For very light $m_{Z'}$, we have $\Delta a_\mu\simeq g^{\prime 2}/8\pi^2$, which requires $g'\sim4\times10^{-4}$ to explain $\Delta a_\mu$. While in the heavy $Z'$ limit, $\Delta a_\mu\simeq g^{\prime 2} m_\mu^2/12\pi^2 m_{Z'}^2$, then the experimental discrepancy needs $g'/m_{Z'}\sim 5\times10^{-3}~\text{GeV}^{-1}$.

For the gauge boson $Z'$, one tight constraint is the dimuon production from the inelastic neutrino-nucleus scattering \cite{Altmannshofer:2014pba}. The CCFR experiment has measured the neutrino trident production cross section, which is consistent with the standard model prediction as $\sigma/\sigma_\text{SM}=0.82\pm0.28$ \cite{CCFR:1991lpl}. This result has excluded the $\Delta a_\mu$ favored region with $m_{Z'}\gtrsim0.3$ GeV. Recently the NA64$\mu$ experiment has excluded $g'\gtrsim6\times10^{-4}$ with $m_{Z'}\lesssim0.1$ GeV \cite{Andreev:2024sgn}. A more stringent constraint comes from the $e^+e^-\to4\mu$ searches at BABAR in the mass region of $0.212<m_{Z'}<10$ GeV \cite{BaBar:2016sci}. For $Z'$ mass in the range of [5, 81] GeV, the LHC experiment also excludes large parameter space by the process $pp\to Z'\mu^+\mu^-\to4\mu$ \cite{ATLAS:2023vxg}.  Meanwhile for $Z'$ lighter than about 6 MeV, it is disfavored by $\Delta N_\text{eff}$ constraints from big-bang nucleosynthesis (BBN) \cite{Escudero:2019gzq}. In the following study, we consider $m_{Z'}\gtrsim 100$ GeV to avoid these tight constraints. 

\subsection{Decay Properties}

Before studying the explicit signatures, we first review the decay properties of new gauge boson $Z'$ and heavy neutral lepton $N$. The partial decay widths of $Z'$ are calculated as
\begin{eqnarray}
	\Gamma(Z'\to \ell^+\ell^-)&=&\frac{g'^2}{12\pi}m_{Z'}, \\
	\Gamma(Z'\to \nu_\ell \nu_\ell)&=&\frac{g'^2}{24\pi}m_{Z'},\\
	\Gamma(Z'\to N_\ell  N_\ell )&=& \frac{g'^2}{24\pi}m_{Z'}\left(1-4\frac{m_{N_\ell }^2}{m_{Z'}^2}\right),
\end{eqnarray}
where $\ell=\mu,\tau$, and we have assumed vanishing masses for $\ell,\nu_\ell$ in the above calculations. The new gauge boson $Z'$ also couples to electron at one loop level. The $Z'\to e^+e^-$ decay width is suppressed by the kinetic mixing factor $\epsilon\simeq g'/70$, thus is not taken into account in this study. 

\begin{figure}
	\begin{center}
		\includegraphics[width=0.45\linewidth]{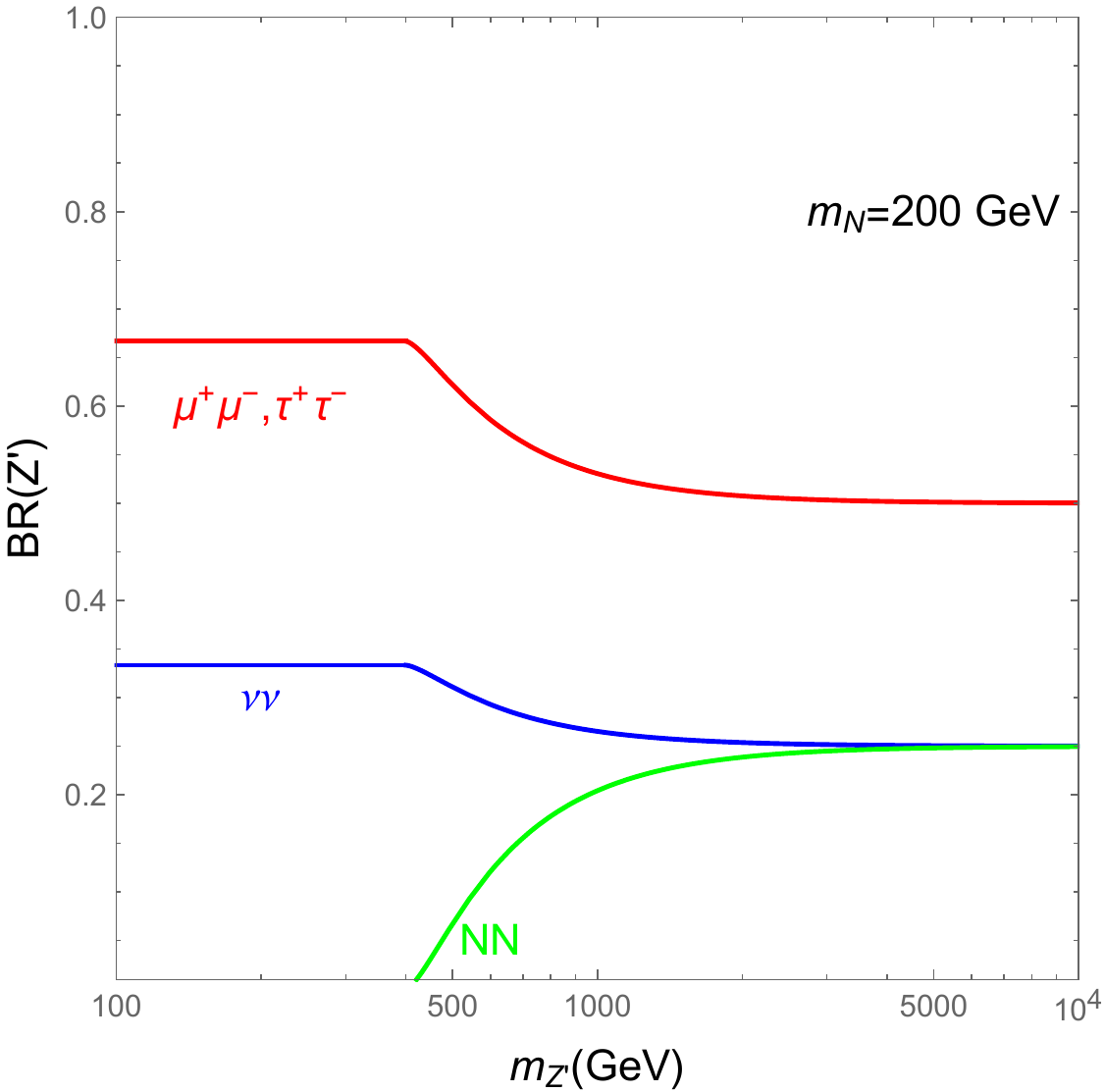}
		\includegraphics[width=0.45\linewidth]{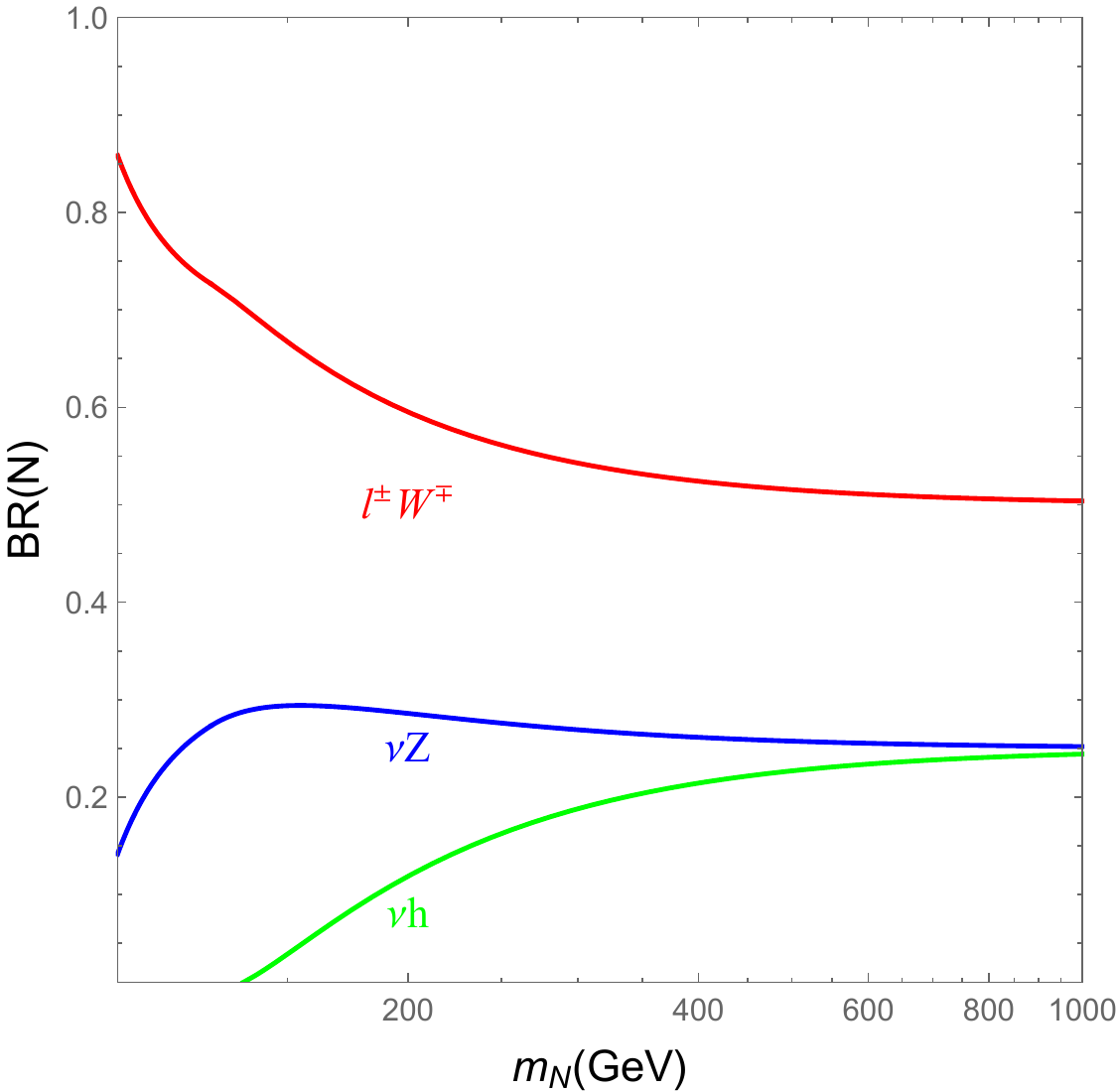}
	\end{center}
	\caption{Branching ratio of gauge boson $Z'$ (left) and heavy neutral lepton $N$ (right).}
	\label{FIG:DC}
\end{figure}

After the production of heavy neutral lepton $N$, it decays via the mixing with light neutrino. When the heavy neutral lepton is heavier than the standard model gauge bosons, the two body decays are the dominant modes.  The partial decay widths are given by  
\begin{eqnarray}
	\Gamma(N\to \ell^\pm W^\mp) & = & \frac{|V_{\ell N}|^2}{16\pi } \frac{(m_N^2-m_W^2)^2(m_N^2+2 m_W^2)}{m_N^3 v_0^2},\\
	\Gamma(N\to \nu Z) & = & \frac{|V_{\ell N}|^2}{32\pi } \frac{(m_N^2-m_Z^2)^2(m_N^2+2 m_Z^2)}{m_N^3 v_0^2},\\
	\Gamma(N\to \nu h) & = & \frac{|V_{\ell N}|^2}{32\pi } \frac{m_N^2-m_h^2}{m_N v_0^2 }.
\end{eqnarray}
In Figure \ref{FIG:DC}, we show the branching ratio (BR) of gauge boson $Z'$ and heavy neutral lepton $N$.  The dilepton mode $Z'\to \mu^+\mu^-,\tau^+\tau^-$ is always the dominant decay channel of $Z'$. Neglecting the final states phase space effect,  we have BR$(Z'\to \nu_\ell \nu_\ell)=$ BR$(Z'\to N_\ell N_\ell )=1/4$ for two degenerate heavy neutral leptons. In the heavy $m_N$ limit, we have BR$(N\to \ell^\pm W^\mp)$: BR$(N\to \nu Z)$: BR$(N\to \nu h)=2:1:1$. In order to reconstruct the heavy neutral lepton mass, we focus on the decay mode $N\to \ell^\pm W^\mp$ with $W\to q \bar{q}'$ in the collider simulation.

\section{Pair Production of Heavy Neutral Lepton}\label{SEC:PR}

Through mixing with light neutrinos, the pair production of heavy neutral leptons at the muon collider  can occur through the $s$-channel $Z$ exchange and $t$-channel $W$ exchange. However, the production cross section is proportional to the fourth power of the mixing parameter $|V_{\ell N}|^4$. Provided natural seesaw predicted value $V_{\ell N}\lesssim10^{-6}$, the predicted cross section is negligible tiny.

The heavy neutral lepton $N$ is charged under the $U(1)_{L_\mu-L_\tau}$ symmetry, which induces the pair production of heavy neutral lepton via $Z'$ at the muon collider. With fixed collision energy, the direct pair production process $\mu^+\mu^-\to NN$ via the $s$-channel off-shell $Z'$ is more promising when $m_{Z'}>\sqrt{s}$. Meanwhile when $m_{Z'}<\sqrt{s}$, the new gauge boson $Z'$ can be produced on-shell with an associated photon from the initial legs as $\mu^+\mu^-\to Z' \gamma \to NN \gamma$, which also requires $m_N<m_{Z'}/2$ to make the decay $Z'\to NN$ kinematically allowed.

\subsection{Without Associated Photon}

We first consider the pair production of heavy neutral leptons without associated photon.
The production cross section of $\mu^+\mu^-\to Z^{\prime *}\to NN$ with a center of mass energy $\sqrt{s}$ is calculated as
\begin{equation}\label{Eq:PR}
	\sigma(\mu^+\mu^-\to Z^{\prime *}\to NN)=\frac{g^{\prime 4}}{24\pi}\frac{s}{(s-m_{Z'}^2)^2+m_{Z'}^2\Gamma_{Z'}^2}\left(1-4\,\frac{m_N^2}{s}\right)^{3/2},
\end{equation}
where $\Gamma_{Z'}$ is the total decay width of $Z'$. 

\begin{figure}
	\begin{center}
		\includegraphics[width=0.45\linewidth]{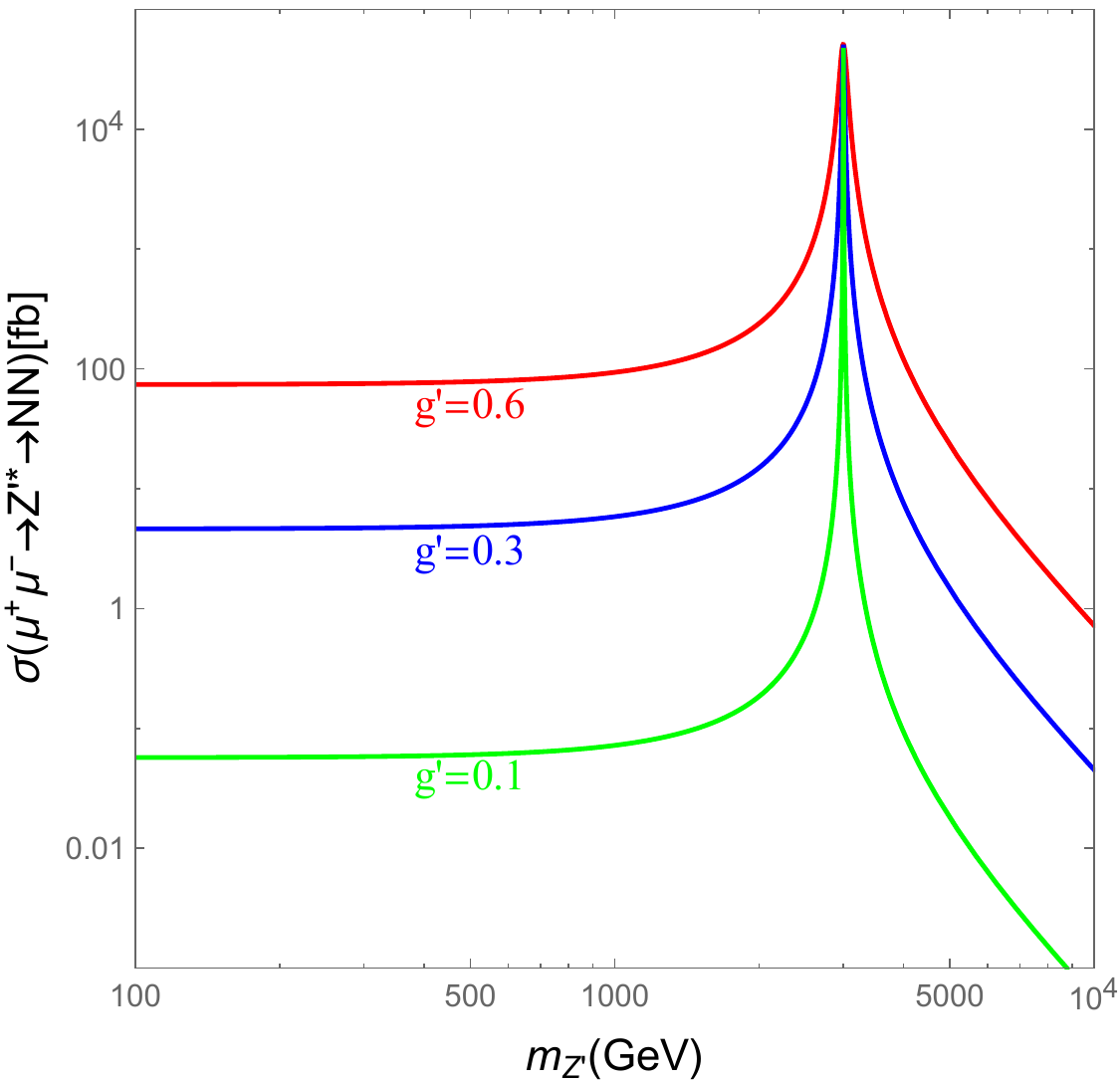}
		\includegraphics[width=0.46\linewidth]{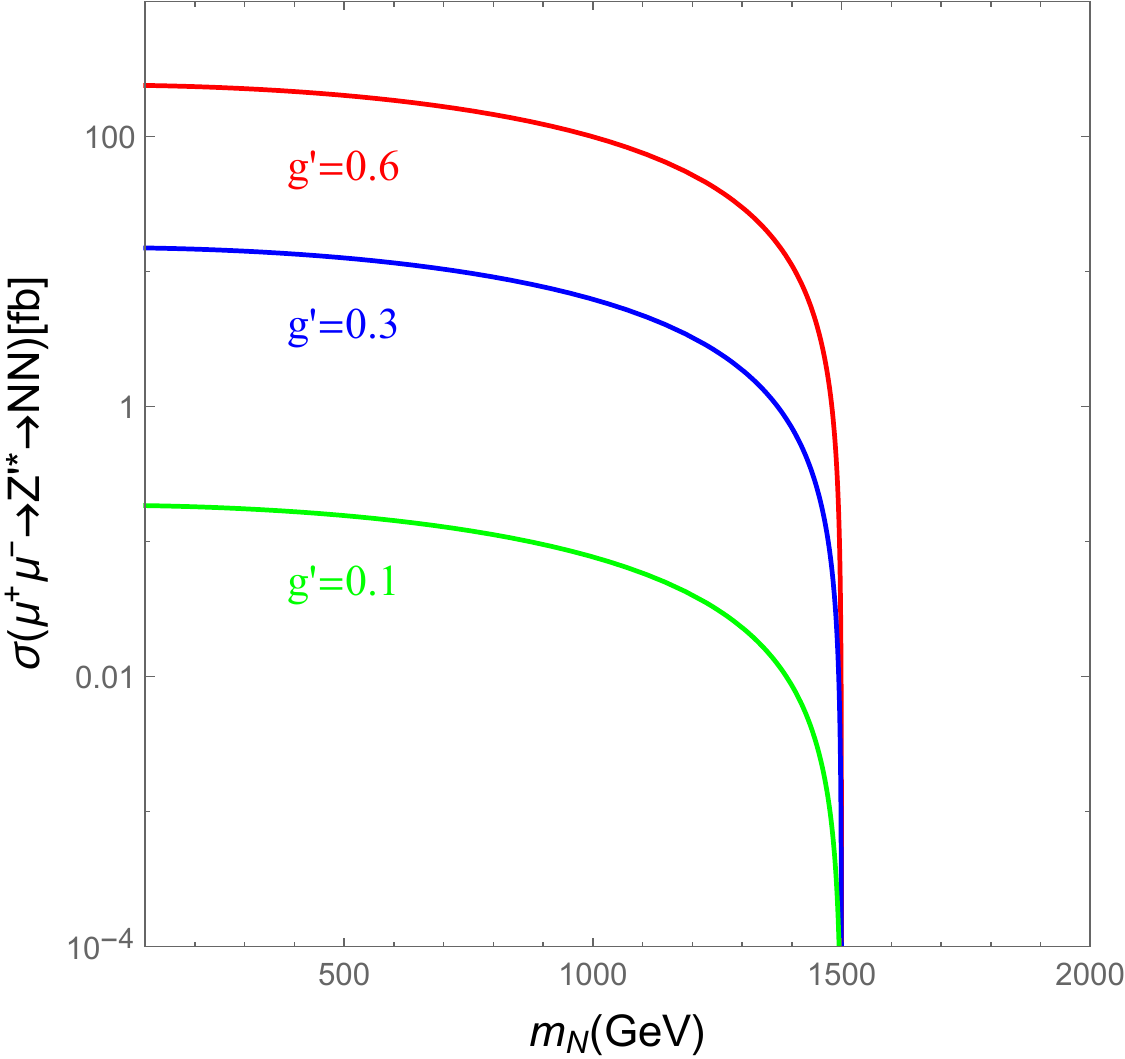}
		\end{center}
	\caption{Cross section of $\mu^+\mu^-\to Z^{\prime *}\to NN$ at a 3 TeV muon collider. We have fixed $m_N=100$~GeV in the left panel and $m_{Z'}=2000$ GeV in the right panel. }
	\label{FIG:ZpNN}
\end{figure}

The theoretical cross section of $\mu^+\mu^-\to Z^{\prime *}\to NN$ at a 3 TeV muon collider for specific scenarios are shown in Figure~\ref{FIG:ZpNN}, where we directly use Eqn~\ref{Eq:PR} without considering the initial state radiation effect. It is obvious that when $m_{Z'}\simeq \sqrt{s}$, the on-shell production of $Z'$ can greatly enhance the cross section. The maximum value is approximately $12\pi \text{BR}(Z'\to \mu^+\mu^-) \text{BR}(Z'\to NN)/m_{Z'}^2\approx 3\pi/(8m_{Z'}^2)$ \cite{Dasgupta:2023zrh}, which could over 50 pb. For a light $Z'$, the cross section reduces to
\begin{equation}
	\sigma(\mu^+\mu^-\to Z^{\prime *}\to NN)\simeq \frac{g^{\prime 4}}{24\pi} \frac{1}{s} \left(1-4\,\frac{m_N^2}{s}\right)^{3/2},
\end{equation}
which is independent of $Z'$ mass. Typically for $g'=0.3$ and $m_N=1000$ GeV, we have $\sigma(\mu^+\mu^-\to Z^{\prime *}\to NN)\simeq4.7$ fb. On the other hand, a heavy $Z'$ leads to 
\begin{equation}
	\sigma(\mu^+\mu^-\to Z^{\prime *}\to NN)\simeq \frac{g^{\prime 4}}{24\pi} \frac{s}{m_{Z'}^4} \left(1-4\,\frac{m_N^2}{s}\right)^{3/2}.
\end{equation}
In this scenario, the cross section is suppressed by the large value of $m_{Z'}$. While single production could probe $m_N\lesssim \sqrt{s}$, the pair production of heavy neutral lepton is kinematically allowed when $m_N<\sqrt{s}/2$. So the 3 TeV muon collider could only probe $m_N\lesssim1500$ GeV. For heavier $m_N$, we need a more energetic muon collider.

\subsection{With Associated Photon}

\begin{figure}
	\begin{center}
		\includegraphics[width=0.45\linewidth]{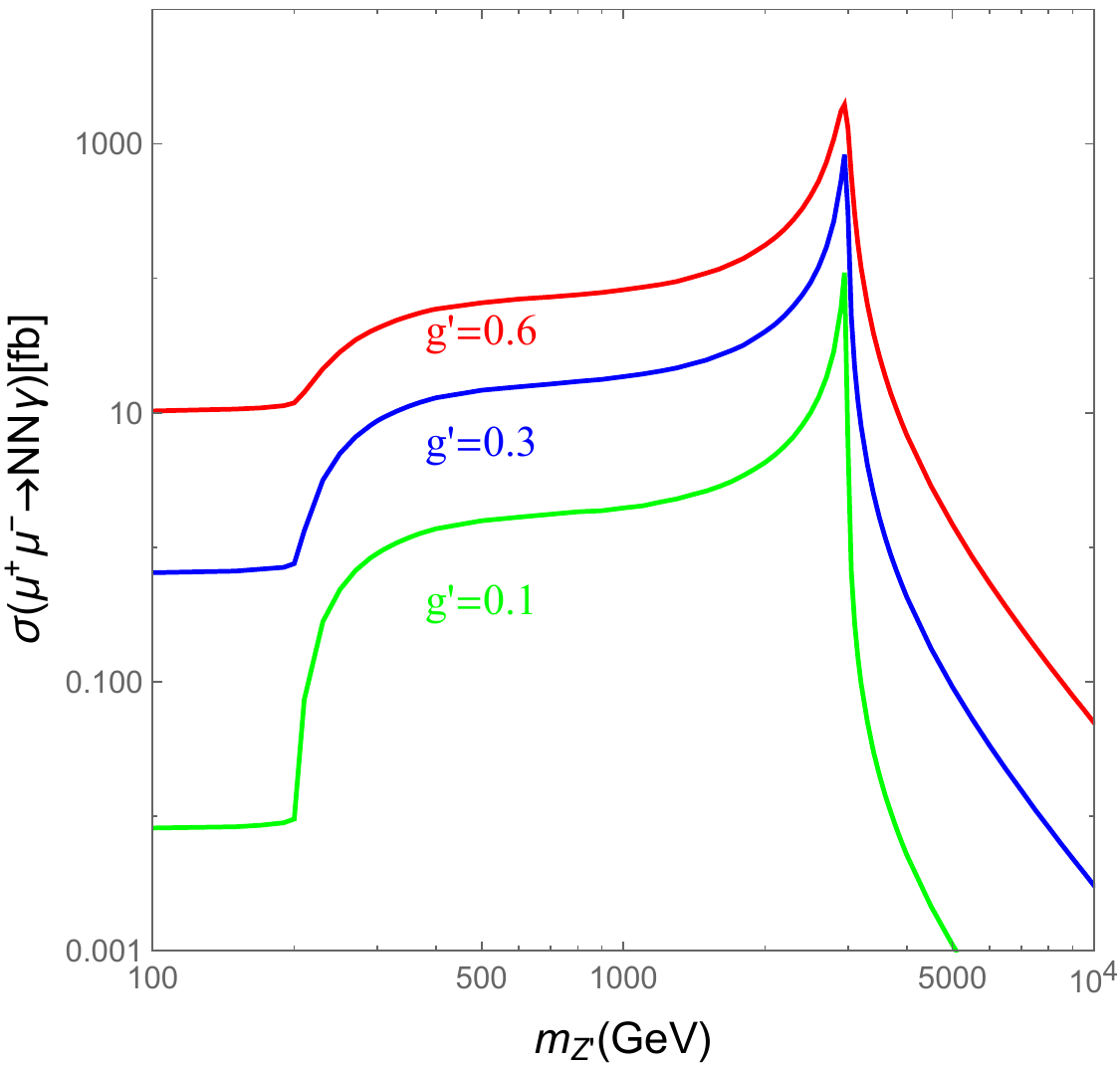}
		\includegraphics[width=0.46\linewidth]{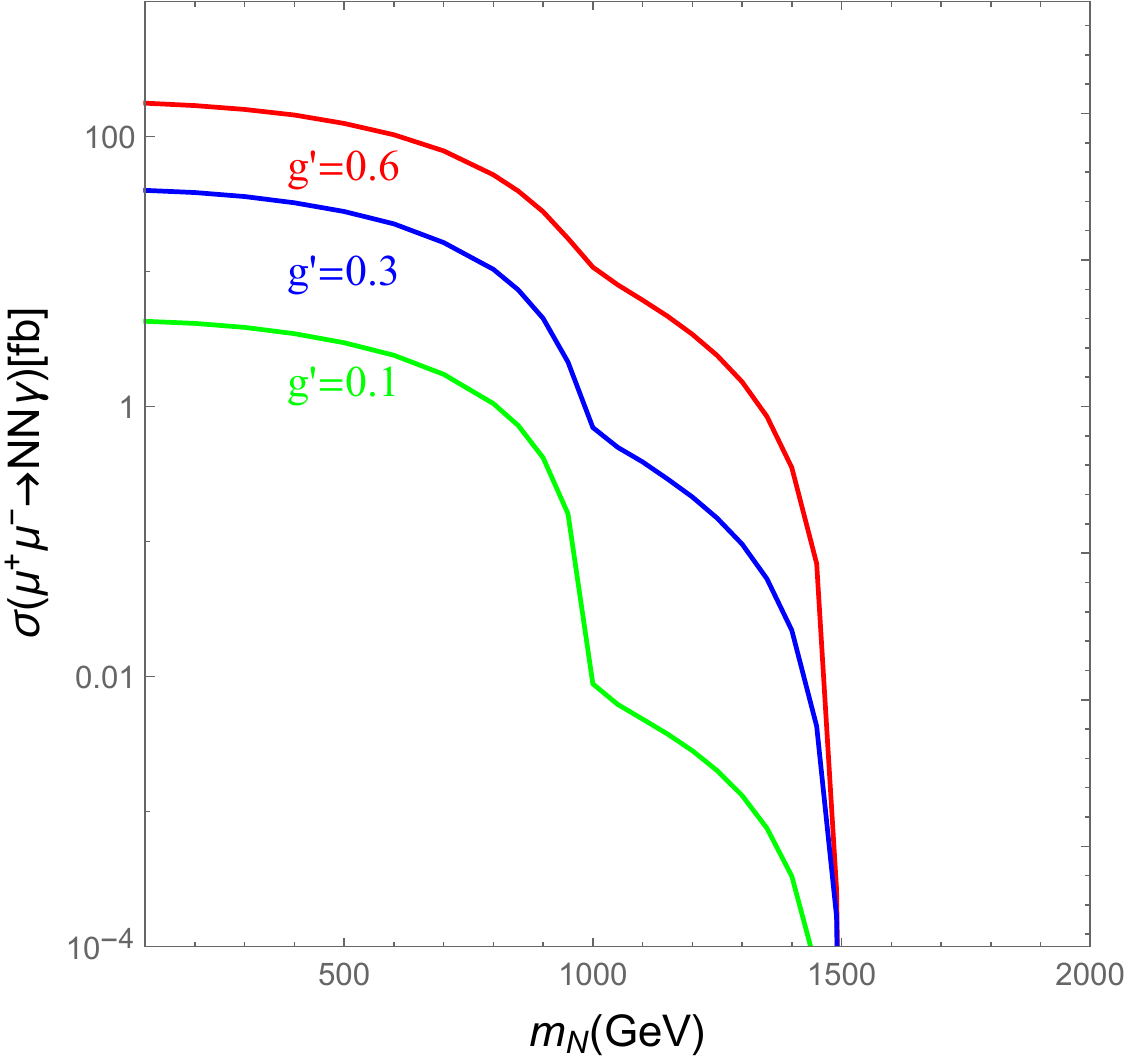}
	\end{center}
	\caption{Cross section of $\mu^+\mu^-\to NN\gamma$ at a 3 TeV muon collider. We have fixed $m_N=100$~GeV in the left panel and $m_{Z'}=2000$ GeV in the right panel. }
	\label{FIG:NNA}
\end{figure}

The photon from initial state radiation carries part of the collision energy, which results in the gauge boson $Z'$ produced on-shell when $m_{Z'}\lesssim \sqrt{s}$. In this case, there is a resonance peak in the invariant mass spectrum of the dilepton \cite{Huang:2021nkl}. Here, we study the pair production of heavy neutral leptons with associated photon process as
\begin{equation}
	\mu^+\mu^-\to Z^{\prime (*)}\gamma\to NN\gamma.
\end{equation}
Here, the scenario with off-shell $Z'$ contribution is also considered when $m_{Z'}\leq 2 m_N$ or $m_{Z'}\geq \sqrt{s}$. 

In Figure \ref{FIG:NNA}, we show the cross section of $\mu^+\mu^-\to NN\gamma$ at a 3 TeV muon collider, where the numerical results are calculated by MadGraph. There is also a sharp peak around $m_{Z'}\simeq \sqrt{s}$. To avoid the soft photon singularity, the detected photon is required to satisfy the following pre-selection cuts
\begin{equation}\label{Eq:cutA}
	P_T(\gamma)>20~\text{GeV}, |\eta(\gamma)|<2.5.
\end{equation}

When $m_{Z'}>2 m_N$, the on-shell production of $Z'$ followed by the cascade decay $Z'\to NN$ can notably enhance the cross section of $\mu^+\mu^-\to NN\gamma$. Using the narrow-width approximation, the cross section can be expressed as \cite{Appelquist:2002mw}
\begin{equation}
	\sigma(\mu^+\mu^-\to NN\gamma)\simeq \sigma(\mu^+\mu^-\to Z'\gamma) \times \text{BR}(Z'\to NN).
\end{equation}
Because the branching ratio of $Z'\to NN$ is independent of the new gauge coupling $g'$, the cross section of $\mu^+\mu^-\to Z'\gamma\to NN\gamma$ is proportional to $e^2g^{\prime 2}$, while the cross section of off-shell process $\mu^+\mu^-\to Z ^{\prime (*)}\gamma\to NN\gamma$ is proportional to $e^2g^{\prime 4}$.
So we observe that the enhancement effect becomes larger when the gauge coupling $g'$ is smaller. For $g'=0.1, m_N=100$ GeV and $m_{Z'}=1000$ GeV, we have $\sigma(\mu^+\mu^-\to Z'\gamma\to NN\gamma)\sim 1$~fb, while $\sigma(\mu^+\mu^-\to Z^{\prime *}\to NN)$ is less than 0.1 fb, therefore the former process is expected more promising at the muon collider.

\section{Lepton Number Violation Signatures}\label{SEC:LNV}

Pair production of heavy neutral lepton followed by various cascade decay modes could lead to many interesting signatures, such as monolepton, dilepton, and trilepton signals \cite{Arun:2022ecj}. In this paper, we focus on the most distinct lepton number violation same-sign dimuon signal, which also has a much cleaner background compared with the lepton number conserving one. 

The simulation procedure is as follows. The FeynRules package \cite{Alloul:2013bka} is used to implement the model. The parton level events are simulated with  MadGraph5\_aMC@NLO \cite{Alwall:2014hca}. We then use PYTHIA 8 \cite{Sjostrand:2014zea} for parton shower. The detector effects are included by employing Delphes 3 \cite{deFavereau:2013fsa} with the detector card of the muon collider. The $W$ boson from the heavy neutral lepton decay could be highly boosted, so the two jets from cascade $W$ decay merge into one fat-jet $J$. We use the Valencia algorithm \cite{Boronat:2014hva} with $R=1.2$ to reconstruct the fat-jets.

\subsection{Without Associated Photon}

\begin{figure}
	\begin{center}
		\includegraphics[width=0.45\linewidth]{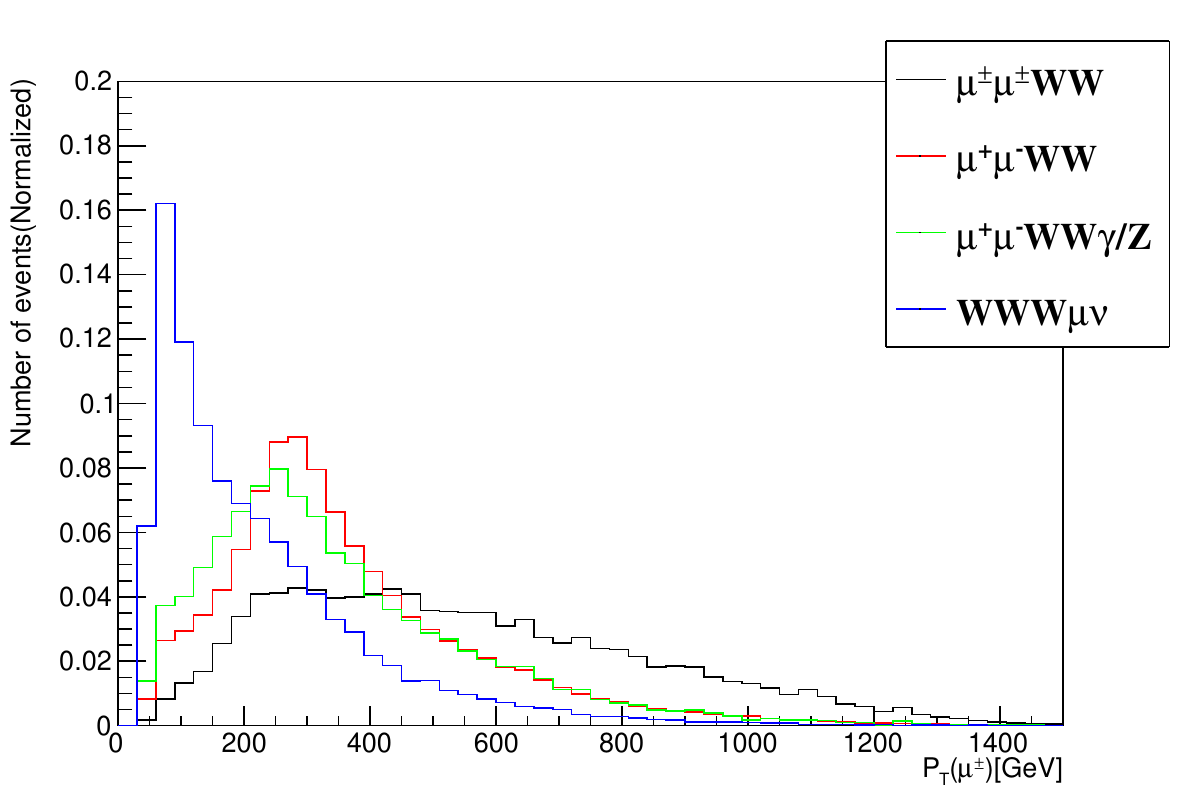}
		\includegraphics[width=0.45\linewidth]{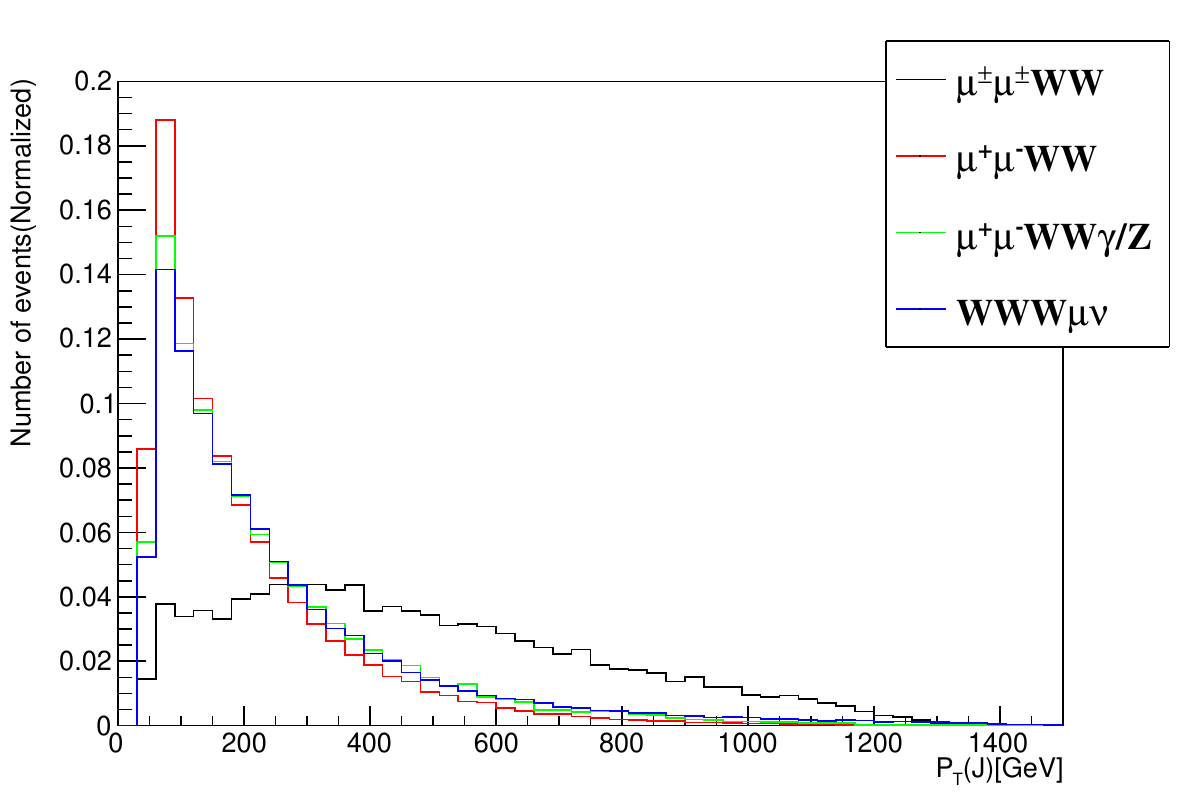}
		\includegraphics[width=0.45\linewidth]{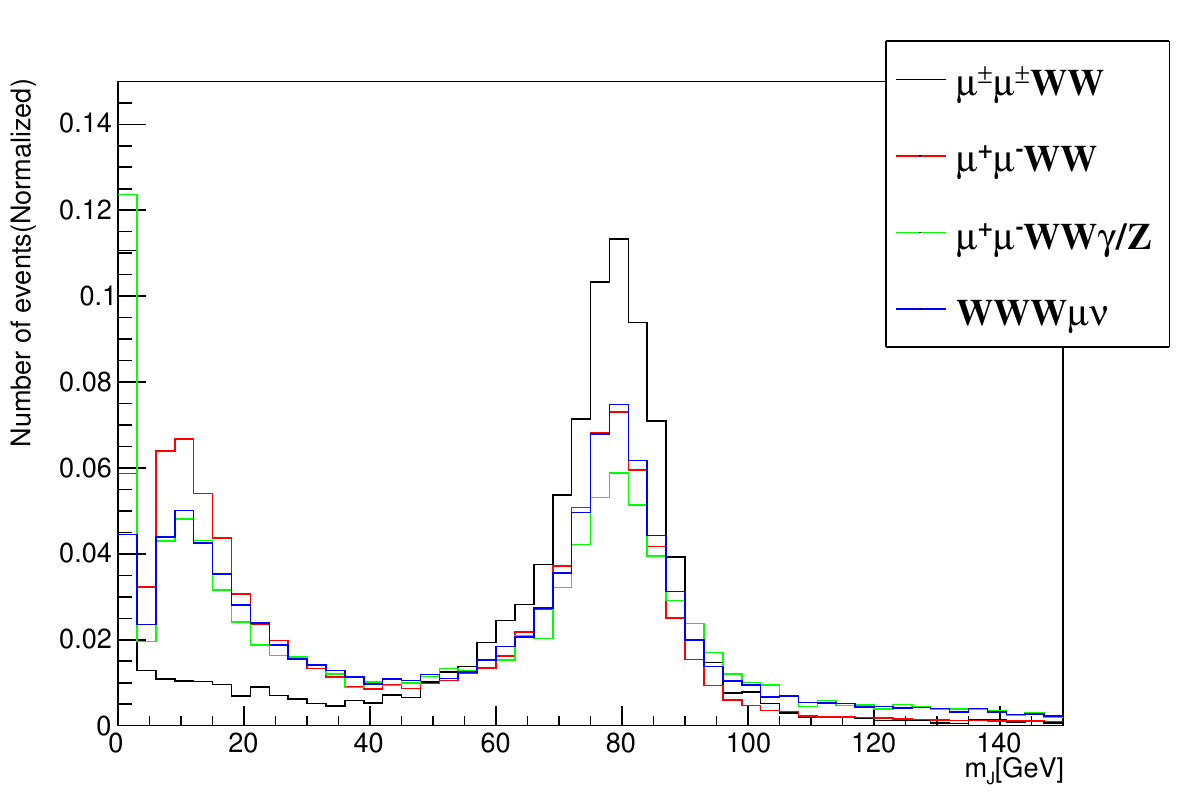}
		\includegraphics[width=0.45\linewidth]{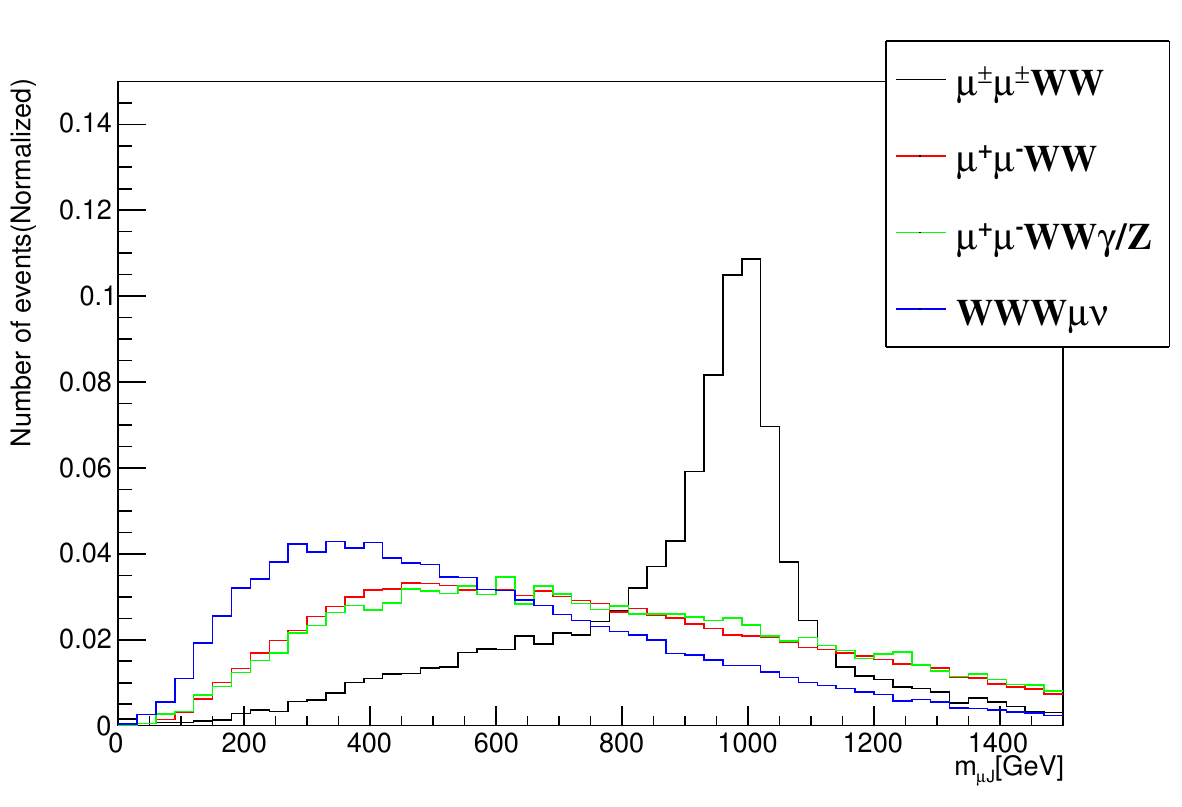}
	\end{center}
	\caption{Normalized distribution of transverse momentum of muon $P_T(\mu^\pm)$ (up-left panel), transverse momentum of fat-jet $P_T(J)$ (up-right panel), fat-jet mass $m_J$ (down-left panel), and invariant mass of muon and fat-jet $m_{\mu J}$ (down-right panel) for the $\mu^\pm\mu^\pm JJ$ signature and corresponding backgrounds.}
	\label{FIG:VA}
\end{figure}

 The full production process of the same-sign dimuon signature without associated photon is
\begin{equation}
	\mu^+\mu^-\to Z^{\prime *}\to N N\to \mu^\pm W^\mp +\mu^\pm W^\mp,
\end{equation}
with the hadronic decays of $W$. The standard model backgrounds come from processes such as
\begin{equation}
	\mu^+\mu^-\to \mu^+\mu^- W^+W^- , \mu^+\mu^-W^+ W^-\gamma/Z,  W^\pm W^\pm W^\mp \mu^\mp \nu.
\end{equation}
The contributions of $\mu^+\mu^- W^+W^-$ and $\mu^+\mu^-W^+ W^-\gamma/Z$ are from lepton charge misidentified, which are suppressed by the misidentification rate  0.1\% \cite{Liu:2021akf}. The $W^\pm W^\pm W^\mp \mu^\mp \nu$ is dominant by the vector boson fusion process, where the two same-sign $W$ bosons decay leptonically. There is also one possible background process $\mu^+\mu^-\to W^+W^-jj$ with the light jets mistagged. However, by requiring the light jets masses close to the $W$ mass, this background can be easily suppressed \cite{Liu:2021akf}.  Meanwhile, the $t\bar{t}W^\pm \mu^\mp\nu$ process also contributes to the same-sign dimuon signature, which can be further reduced by cut on the opposite-sign lepton \cite{Li:2023lkl}. So we do not include contributions of the $\mu^+\mu^-\to W^+W^- jj$ and $t\bar{t}W^\pm \mu^\mp\nu$ processes in this paper.

We first apply the following pre-selection cuts on the transverse momentum and pseudorapidity of the muon and the fat-jets
\begin{equation}
	P_T (\mu^\pm)>50~\text{GeV}, |\eta (\mu^\pm)|<2.5,P_T (J)>50~\text{GeV}, |\eta (J)|<2.5.
\end{equation}
In Figure~\ref{FIG:VA}, we show the distributions of some variables for the $\mu^\pm\mu^\pm JJ$ signal and backgrounds after applying the pre-selection cuts.  We have set $m_{N}=1000$~GeV, $m_{Z'}=2500$~GeV and $g'=0.6$ as the benchmark point of the signal. The cut flow for the  $\mu^\pm\mu^\pm JJ$ signature and backgrounds are summarized in Table~\ref{Tab:NmNm}.

The lepton number violation signature is satisfied for events with two same-sign muons.
In the single production of the heavy neutral lepton process \cite{Li:2023lkl,Antonov:2023otp}, only one boosted $W$ is expected from $N$ decay. To distinguish from the single production and also suppress the background from vector boson fusion processes, we require exact two fat-jets in the final states 
\begin{equation}
	N_{\mu^\pm}=2,	~N_J=2,
\end{equation}
although tagging one fat-jet has a larger efficiency for the signal.
In order to be identified as $W$ bosons, the fat-jets masses are also required in the following range  
\begin{equation}
	50 ~\text{GeV} \leq m_J \leq 100 ~\text{GeV}.
\end{equation}

Because there are two heavy neutral leptons in the signal, we reconstruct their masses through the two muons and two fat-jets system by minimizing $\chi^2=(m_{\mu_1 J_1} -m_N)^2+(m_{\mu_2 J_2} -m_N)^2$ \cite{Liu:2021akf}. We require that both the invariant mass of muon and fat-jet $m_{\mu J}$ from reconstructed $N$ satisfy
\begin{equation}
	0.8 m_N<m_{\mu J}<1.2 m_N.
\end{equation}

\begin{table}
	\begin{center}
		\resizebox{\linewidth}{!}{
			\begin{tabular}{c| c | c | c |c } 
				\hline
				\hline
				$\sigma$(fb) & $\mu^\pm\mu^\pm W^\mp W^\mp$ & $\mu^+\mu^- W^+W^-$ & $\mu^+\mu^-W^+ W^-\gamma/Z$  & $W^\pm W^\pm W^\mp \mu^\mp \nu$  \\
				\hline
				Pre-selection	 & 68.19   &   21.57  & 2.13 &  3.11  \\ 
				\hline
				$N_{\mu^\pm}=2$ &  41.38  & $2.6\times10^{-2}$  & $2.5\times10^{-3}$  &   $1.8\times10^{-1}$   \\ 
				\hline
				$N_J=2$ & 24.87  &  $1.1\times10^{-2}$   & $1.3\times10^{-3}$  &   $7.1\times10^{-2}$  \\ 
				\hline
				50 GeV $\leq m_J\leq$ 100 GeV & 14.41 &  $3.2\times10^{-3}$   & $2.1\times10^{-4}$  &   $2.0\times10^{-2}$  \\
				\hline
				$0.8 m_N<m_{\mu J}<1.2 m_N$& 13.62  &  $3.4\times10^{-4}$   & $2.7\times10^{-5}$  &   $8.1\times10^{-4}$   \\
				\hline \hline
			\end{tabular}
		}
	\end{center}
	\caption{Cut flow table for the $\mu^\pm\mu^\pm JJ$ signal at the 3 TeV muon collider and corresponding backgrounds.
		\label{Tab:NmNm}}
\end{table} 

\begin{figure}
	\begin{center}
		\includegraphics[width=0.45\linewidth]{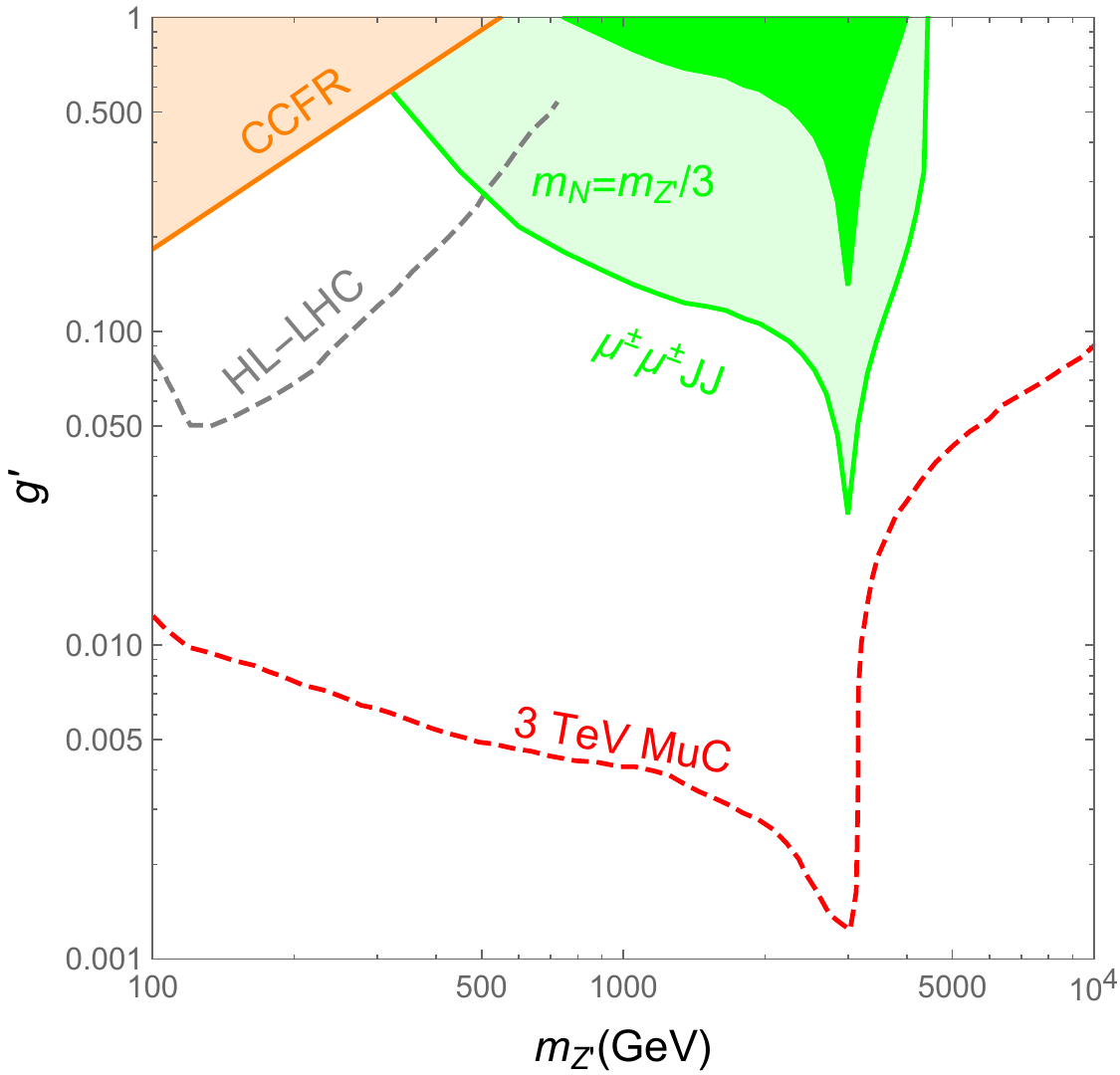}
		\includegraphics[width=0.45\linewidth]{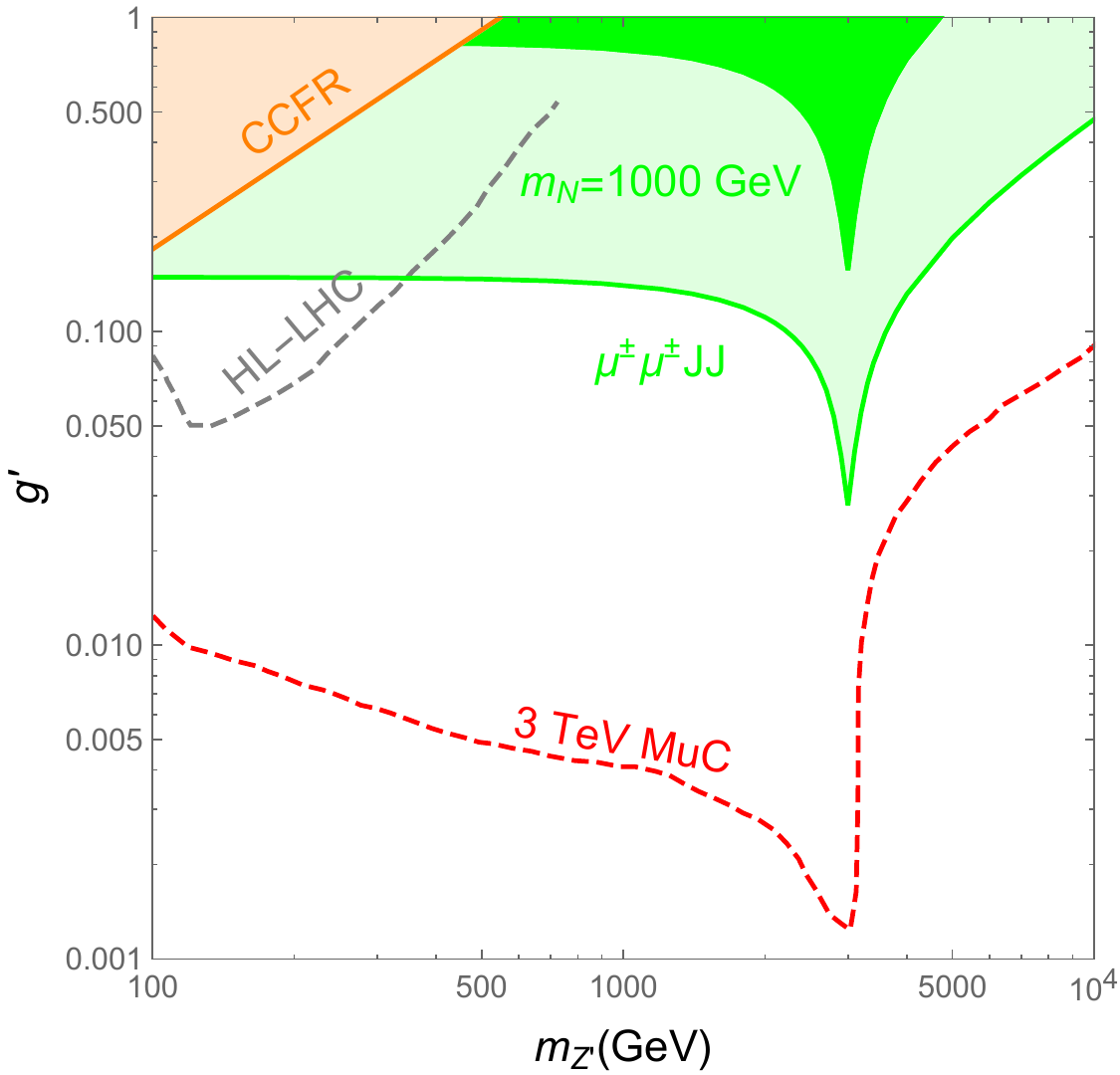}
		\includegraphics[width=0.45\linewidth]{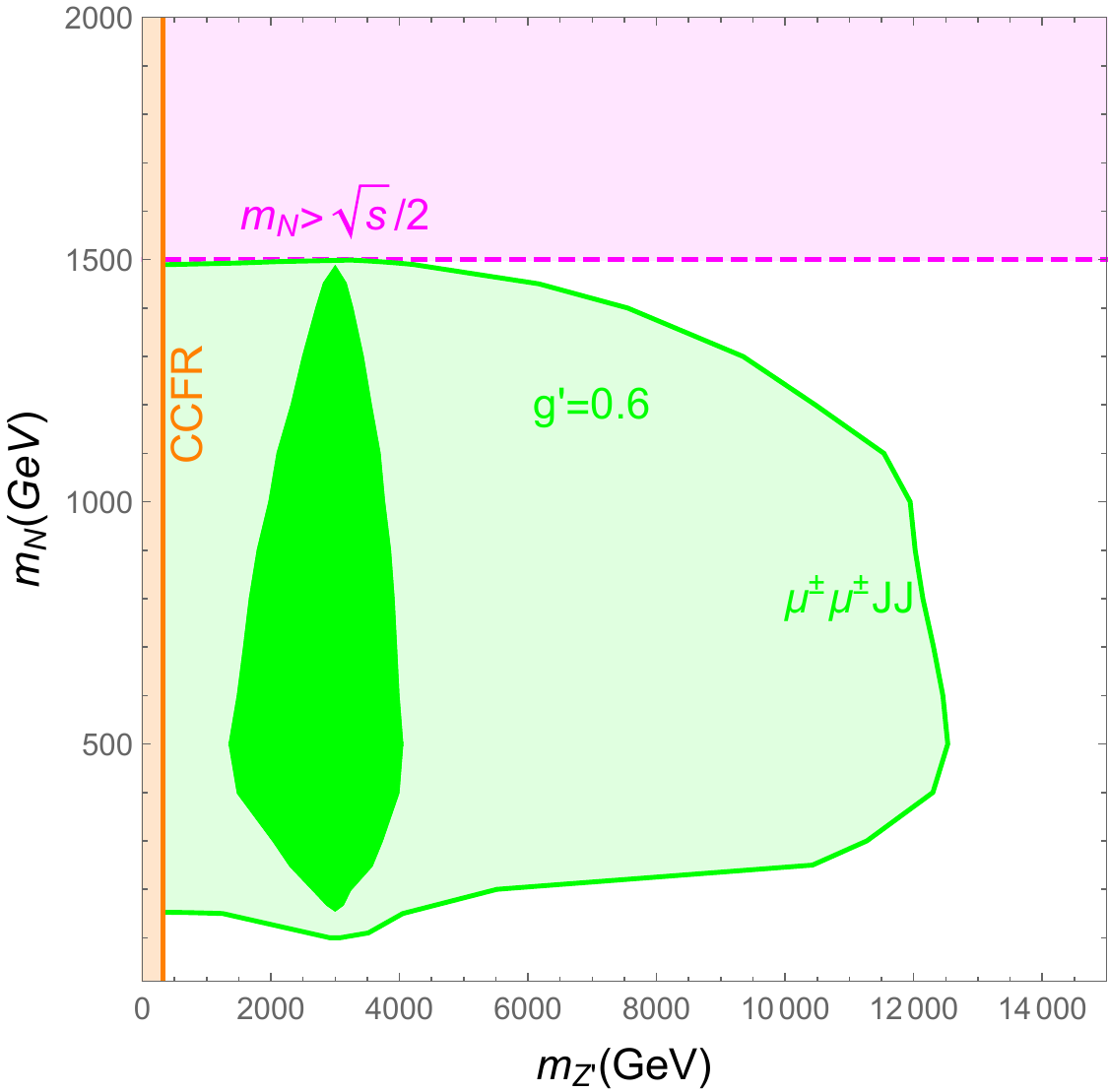}
	\end{center}
	\caption{Sensitivity of the $\mu^\pm \mu^\pm JJ$ signature at the 3 TeV muon collider. In the up-left panel, we fix the mass relation with $m_N=m_{Z'}/3$. In the up-right panel, we set $m_N=1000$ GeV. In the down panel, we fix the gauge coupling $g'=0.6$. The dark-green region corresponds to sensitivity with 1 fb $^{-1}$ data, while the light-green region corresponds to sensitivity with 1000 fb$^{-1}$ data. The orange region is excluded by neutrino trident production at CCFR \cite{CCFR:1991lpl}. The magenta region is not allowed kinematically. The gray line is the projected sensitivity at HL-LHC \cite{delAguila:2014soa}. The red line is the sensitivity of the 3 TeV muon collider \cite{Huang:2021nkl}. }
	\label{FIG:SC1}
\end{figure}

According to the distributions in Figure~\ref{FIG:VA}, we may tighten the cuts on $P_T(\mu^\pm)$ and $P_T(J)$ to suppress the background. However, we see from Table~\ref{Tab:NmNm} that the cut on the invariant mass $m_{\mu J}$ is quite efficient to reduce the background. After applying all these cuts, the total cross section of the background is about $1.2\times10^{-3}$ fb. So with an integrated luminosity of 1000 fb$^{-1}$, there are only 1.2 background events, while the signal events is over $10^4$. In this way, the significance will be 116.7 for the benchmark point with 1000 fb$^{-1}$ data, where the significance is calculated as
\begin{equation}
	\mathcal{S}=\frac{N_S}{\sqrt{N_S+N_B}}.
\end{equation}
Here, $N_S$ and $N_B$ are the event number of signal and background respectively. To reach the $5\sigma$ discovery limit, we need only 1.8 fb$^{-1}$ data.

Based on the above analysis, we then explore the sensitivity of the $\mu^\pm \mu^\pm JJ$ signature at the 3 TeV muon collider. The results are shown in Figure~\ref{FIG:SC1}. There are three free parameters as $g',m_{Z'}$ and $m_N$ related to the collider phenomenology. With fixed mass relation $m_{N}=m_{Z'}/3$, we could probe $m_{Z'}$ approximately in the range of [330, 4500]~GeV. Around the resonance region $m_{Z'}\sim3000$ GeV, the gauge coupling $g'$ could be down to about 0.03 with 1000 fb$^{-1}$ data. We also find that for $m_N\lesssim300$ GeV, the cut efficiency decreases quickly as $m_N$ becomes smaller. One main reason is that the heavy neutral leptons are also highly boosted at the 3 TeV muon collider for such light $m_N$, so the muons and $W$-jets from boosted $N$ decays are mostly non-isolated \cite{Dey:2022tbp}. One may use the substructure based variables as lepton sub-jet fraction (LSF) and lepton mass drop (LMD) \cite{Brust:2014gia} to probe the light $m_N$, which is beyond the scope of this study.

For pair production of heavy neutral leptons without associated photon signature, the gauge boson $Z'$ does not need to be heavier than $2 m_N$. In the up-right panel of Figure~\ref{FIG:SC1}, we have fixed $m_N=1000$ GeV for illustration. The production cross section of $\mu^+\mu^-\to NN$ is approximately a constant for light $Z'$, so $g'\sim0.15$ is nearly independent of $m_{Z'}$ when $Z'$ is lighter than 1 TeV. This channel is most sensitive for large $m_{Z'}$ region. For instance, we may probe $m_{Z'}\sim10$ TeV with $g'\gtrsim0.5$ and $m_N\sim1000$ GeV at the 3~TeV muon collider. Near the resonance region with $m_{Z'}\sim \sqrt{s}$, we even can have a promising signature when $g'\gtrsim0.2$ with only 1 fb$^{-1}$ data.

We then fix the gauge coupling $g'=0.6$ and explore the sensitivity region on the $m_N-m_{Z'}$ plane. With 1 fb$^{-1}$ data, a large parameter space near the resonance region with 2 TeV $\lesssim m_{Z'}\lesssim4$ TeV can be probed. The most sensitive heavy neutral lepton mass is around 500~GeV. Because a lighter $m_N$ leads to smaller acceptance efficiency, while a larger $m_N$ has a smaller production cross section. With an integrated luminosity of 1000~fb$^{-1}$, the region with $m_{Z'}\lesssim 12$ TeV and 150 GeV $\lesssim m_N<1500$ GeV is within the reach of the 3 TeV muon collider.

\subsection{With Associated Photon}

\begin{table}
	\begin{center}
		\begin{tabular}{c| c | c |c } 
			\hline
			\hline
			$\sigma$(fb) & $\mu^\pm\mu^\pm W^\mp W^\mp \gamma$ & $\mu^+\mu^- W^+W^- \gamma$   & $W^\pm W^\pm W^\mp\gamma \mu^\mp \nu$  \\
			\hline
			Pre-selection	 & 24.91   &  1.15   &  0.14    \\ 
			\hline
			$N_{\gamma}=1,N_{\mu^\pm}=2$ & 13.28   & $1.1\times10^{-3}$  &  $7.4\times10^{-3}$     \\ 
			\hline
			$N_J\geq1$ & 13.25 & $1.0\times10^{-3}$    &    $7.2\times10^{-3}$     \\ 
			\hline
			50 GeV $\leq m_J\leq$ 100 GeV & ~~9.61 & $5.7\times10^{-4}$    &  $4.5\times10^{-3}$     \\
			\hline
			$0.8 m_N<m_{\mu J}<1.2 m_N$& ~~8.86  &  $2.6\times10^{-4}$   &    $1.2\times10^{-3}$      \\
			\hline \hline
		\end{tabular}
	\end{center}
	\caption{Cut flow table for the $\mu^\pm\mu^\pm JJ\gamma$ signal at the 3 TeV muon collider and corresponding backgrounds.
		\label{Tab:NmA}}
\end{table} 

The full production process of the same-sign dimuon signature with associated photon is
\begin{equation}
	\mu^+\mu^-\to Z^{\prime (*)} \gamma \to N N \gamma\to \mu^\pm W^\mp +\mu^\pm W^\mp+\gamma,
\end{equation}
with the hadronic decays of $W$. The standard model backgrounds come from processes such as
\begin{equation}
	\mu^+\mu^-\to \mu^+\mu^- W^+W^-\gamma ,  W^\pm W^\pm W^\mp \gamma \mu^\mp \nu .
\end{equation}
During the simulation, the pre-selection cuts on associated photon in Eqn~\eqref{Eq:cutA} are also applied. Based on the cut-flow in Table \ref{Tab:NmNm}, we expect that the cross section of the process $\mu^+\mu^-W^+ W^-\gamma\gamma/Z$ with one additional $\gamma/Z$ gauge boson is about an order of magnitude smaller than the process $\mu^+\mu^- W^+W^-\gamma$. Therefore, $\sigma(\mu^+\mu^-W^+ W^-\gamma\gamma/Z)$ is approximately $10^{-5}$ fb after all cuts, which will not be considered in the analysis of the $\mu^\pm\mu^\pm JJ\gamma$ signal.

\begin{figure}
	\begin{center}
		\includegraphics[width=0.45\linewidth]{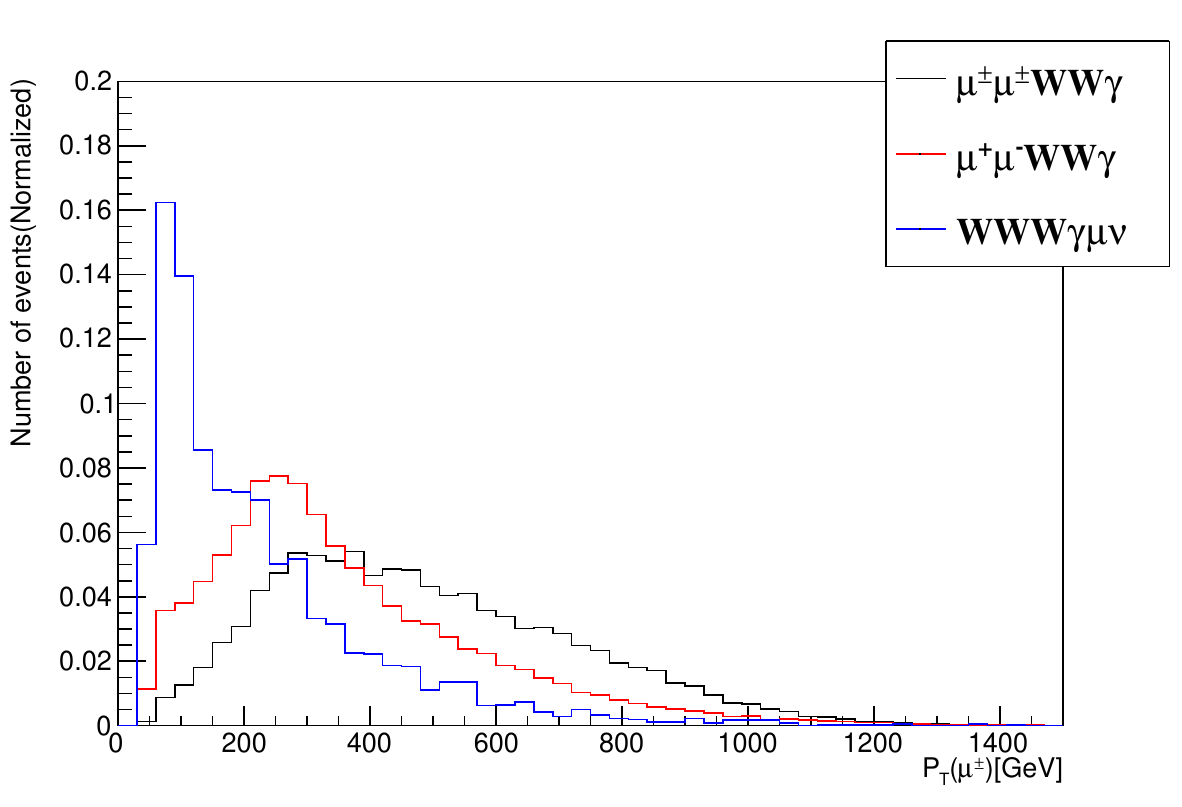}
		\includegraphics[width=0.45\linewidth]{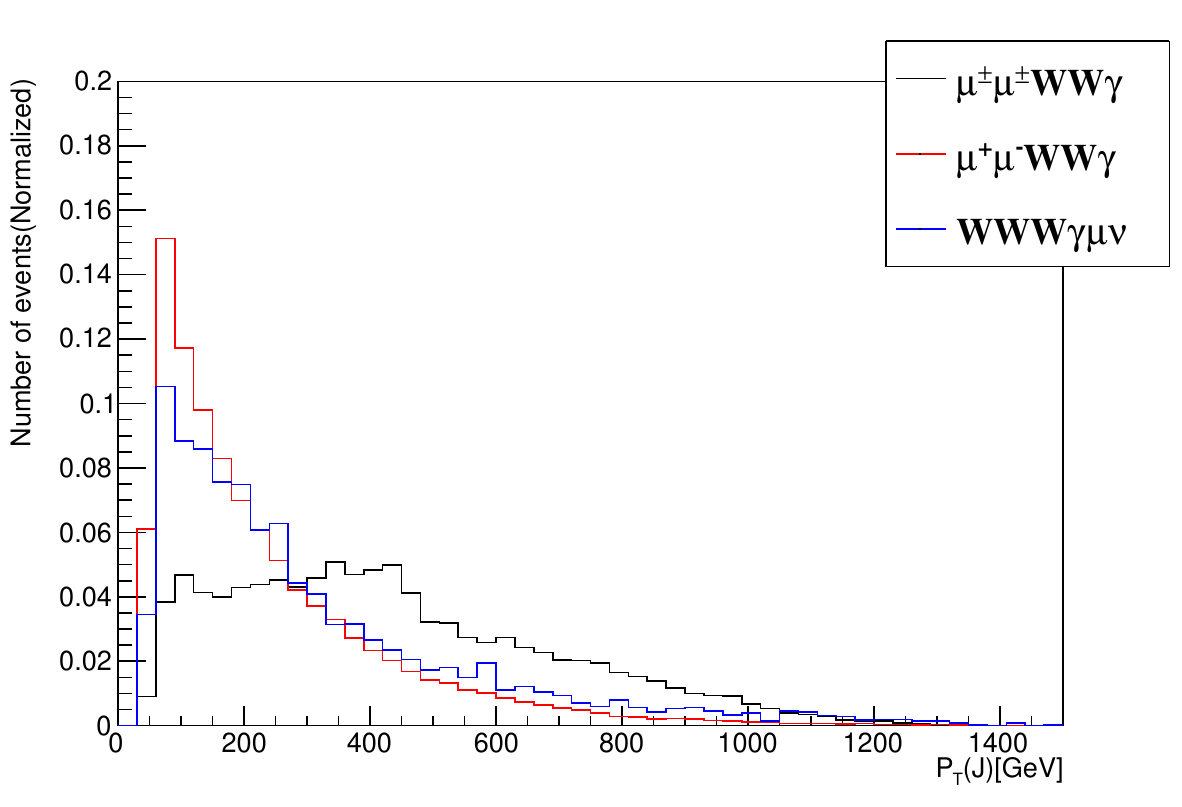}
		\includegraphics[width=0.45\linewidth]{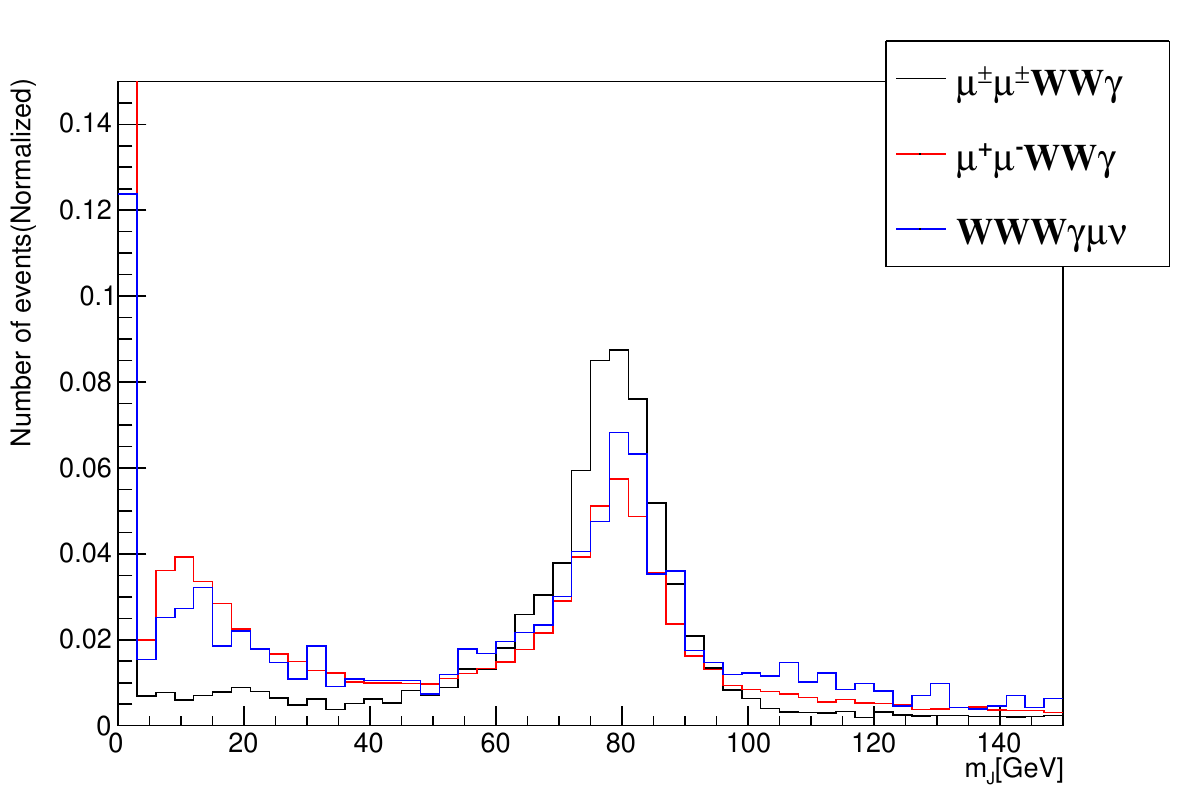}
		\includegraphics[width=0.45\linewidth]{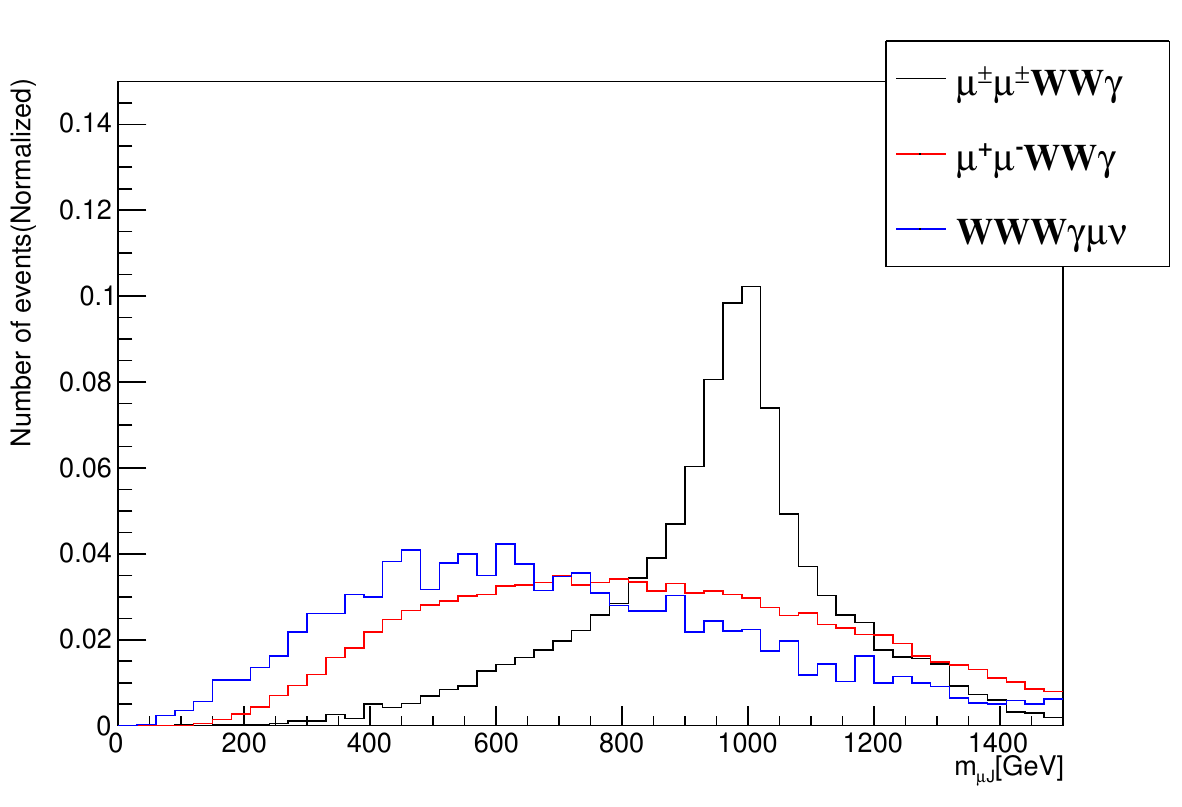}
		\includegraphics[width=0.45\linewidth]{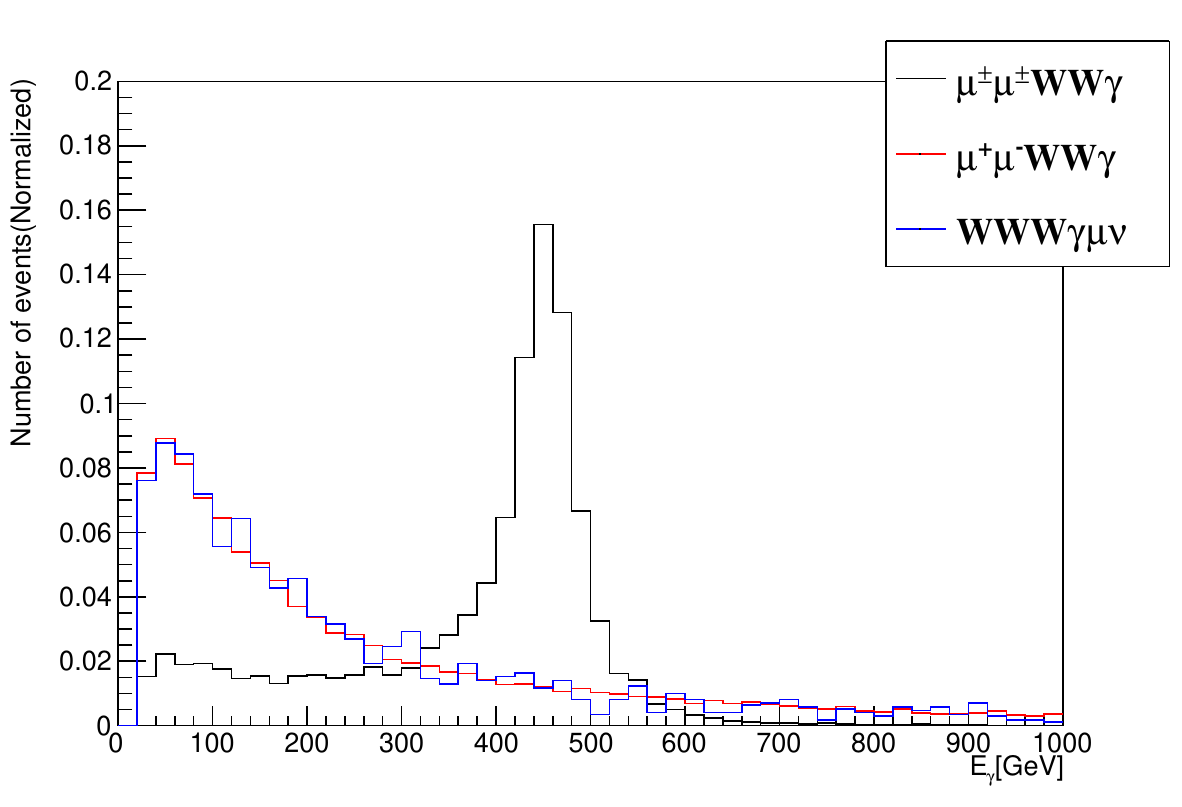}
		\includegraphics[width=0.45\linewidth]{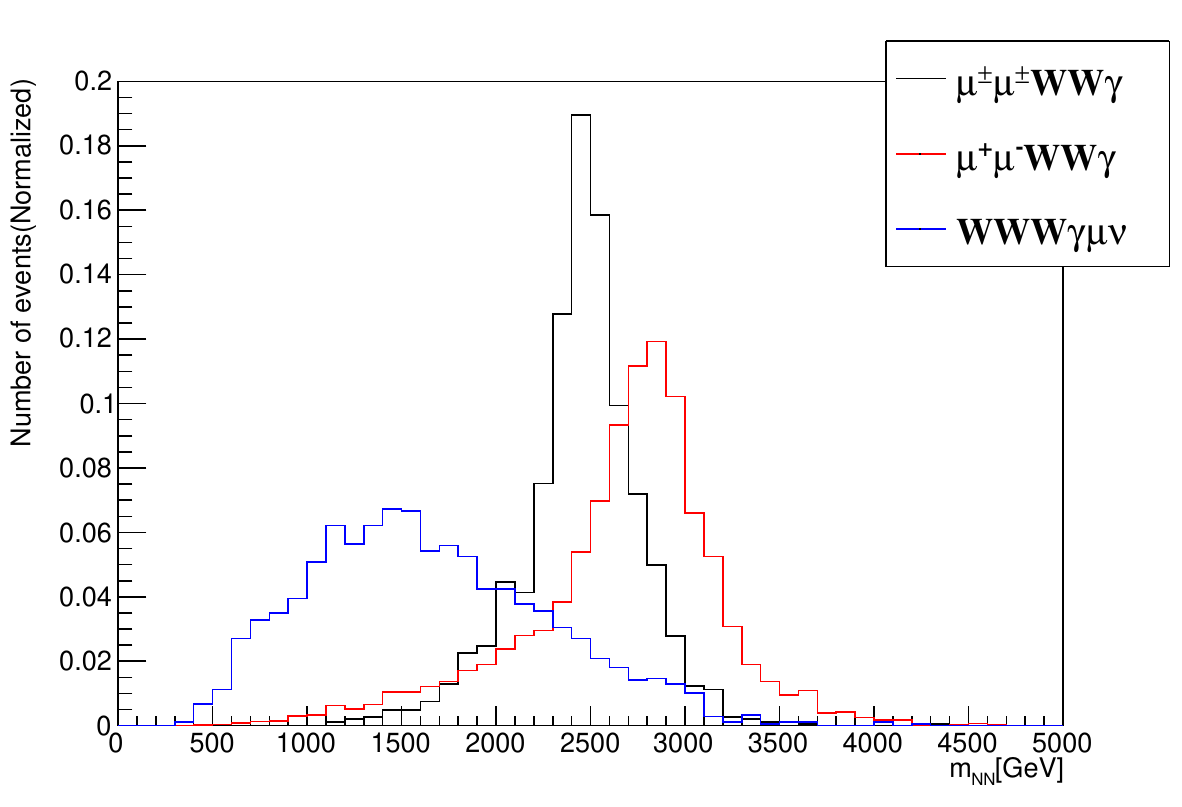}
	\end{center}
	\caption{Normalized distribution of transverse momentum of muon $P_T(\mu^\pm)$ (up-left panel), transverse momentum of fat-jet $P_T(J)$ (up-right panel), fat-jet mass $m_J$ (middle-left panel), and invariant mass of muon and fat-jet $m_{\mu J}$ (middle-right panel), energy of photon $E_\gamma$ (down-left panel) and invariant mass of reconstructed heavy neutral lepton pair $m_{NN}$ (down-right panel)  for the $\mu^\pm\mu^\pm JJ \gamma$ signature and corresponding backgrounds.}
	\label{FIG:VB}
\end{figure}
The cross sections of the background processes with associated photons are about one order of magnitude smaller than those without associated photons. If we apply exactly the same selection cuts as the previous $\mu^\pm\mu^\pm JJ$ signal, the total cross section of background is expected at the order of $10^{-4}$ fb, which will have about 0.1 background event even with a total of 1000 fb$^{-1}$ data. Therefore, we can loosen the selection cut to keep more signal events. In the analysis of the $\mu^\pm\mu^\pm JJ\gamma$ signal, we consider that at least one fat-jet is detected in the final states
\begin{equation}
	N_J\geq 1,
\end{equation}
while we further require one detected photon in this signal.

In Figure \ref{FIG:VB}, we show the distributions of some variables for the $\mu^\pm\mu^\pm JJ\gamma$ signal and backgrounds after applying the pre-selection cuts. The benchmark point for the $\mu^\pm\mu^\pm JJ\gamma$ signal is the same as the previous $\mu^\pm\mu^\pm JJ$ signal.  With quite similar distributions of $P_T(\mu^\pm)$, $P_T(J)$, $m_J$ and $m_{\mu J}$, we then apply the same selection cuts for these variables as the $\mu^\pm\mu^\pm JJ$ signal.  The cut flow for the  $\mu^\pm\mu^\pm JJ\gamma$ signature and backgrounds are summarized in Table~\ref{Tab:NmA}.

Although with one additional photon, the cross section of the $\mu^\pm\mu^\pm JJ\gamma$ signal is in the same order as the  $\mu^\pm\mu^\pm JJ$ signal. After all selection cuts, the cross section of the $\mu^\pm\mu^\pm JJ\gamma$ process is 8.86 fb. With a loose cut on the fat-jet number,  the total cross section of the background is about $1.5\times10^{-3}$~fb after all cuts. With an integrated luminosity of 1000~fb$^{-1}$, the significance of the  $\mu^\pm\mu^\pm JJ\gamma$ signal could be over 90 for the benchmark point. Meanwhile, 2.9 fb$^{-1}$ data is enough to discover the benchmark point at the $5\sigma$ level.

Besides the invariant mass of dilepton $m_{\ell^+\ell^-}(\ell=\mu,\tau)$, there are two more pathways to confirm the on-shell production of gauge boson $Z'$ in the $\mu^\pm \mu^\pm JJ\gamma$ signature. For the two-body process $\mu^+\mu^-\to Z' \gamma$, the photon energy is connected to the gauge boson mass as \cite{Huang:2021nkl}
\begin{equation}
	E_\gamma=\frac{s-m_{Z'}^2}{2\sqrt{s}}.
\end{equation}
For the benchmark point with $m_{Z'}=2500$ GeV at the 3 TeV muon collider, the peak value of $E_\gamma$ is about 458 GeV. The sharp bump above the background is clearly shown in the down-left panel of Figure~\ref{FIG:VB}. On the other hand, we can also measure $m_{Z'}$ through the invariant mass of two reconstructed heavy neutral leptons $m_{NN}$, which also has a resonance near $m_{Z'}$. From the distribution of $m_{NN}$ in the down-right panel of Figure~\ref{FIG:VB}, the background process $\mu^+\mu^-W^+W^-\gamma$ and $W^\pm W^\pm W^\mp \gamma \mu^\mp \nu$ have peaks around 3 TeV and 1.5 TeV respectively. Anyway, with less than two background events after all cuts, the backgrounds have little impact on the distribution of $m_{NN}$ for the signal.

\begin{figure}
	\begin{center}
		\includegraphics[width=0.45\linewidth]{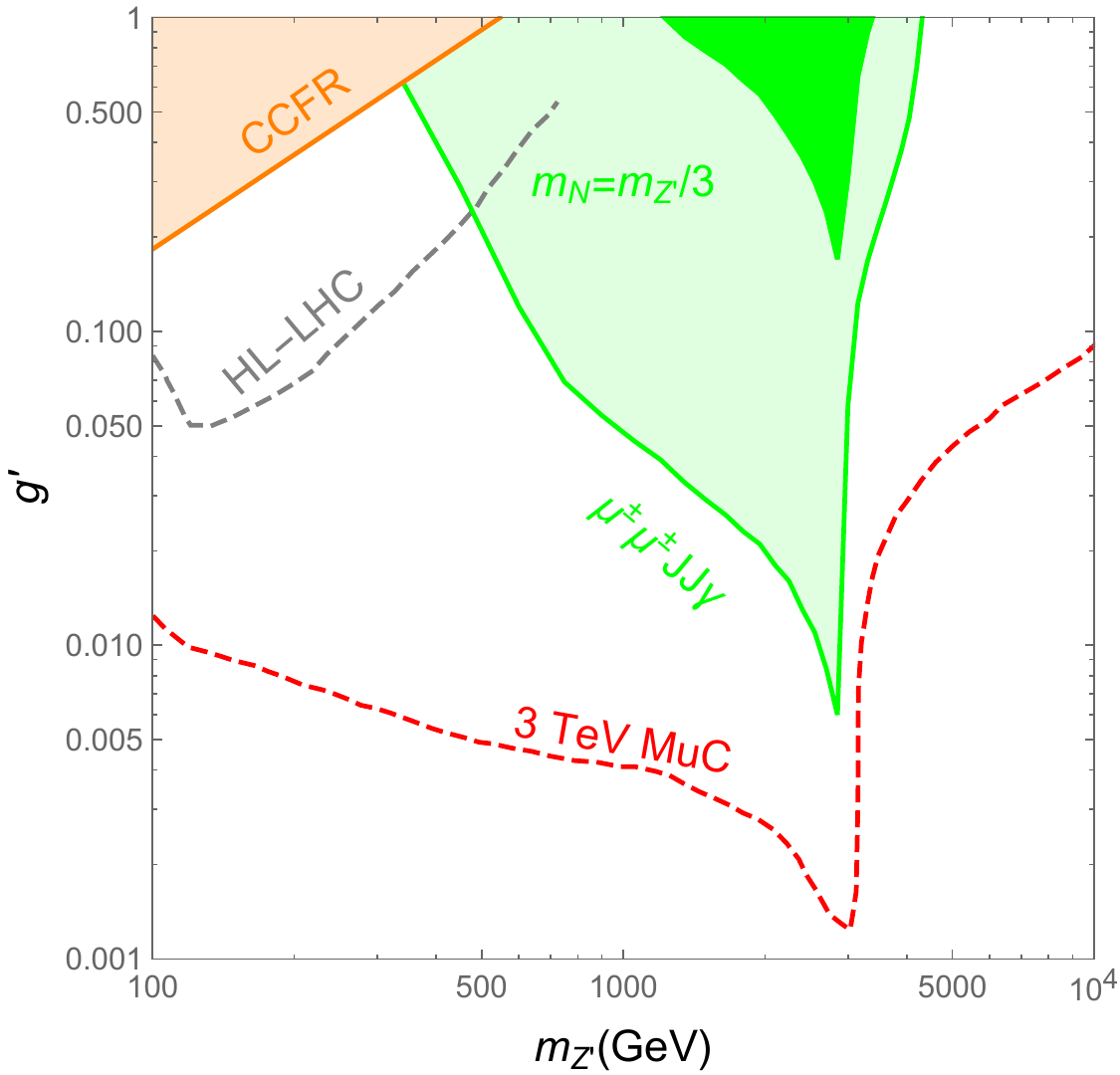}
		\includegraphics[width=0.45\linewidth]{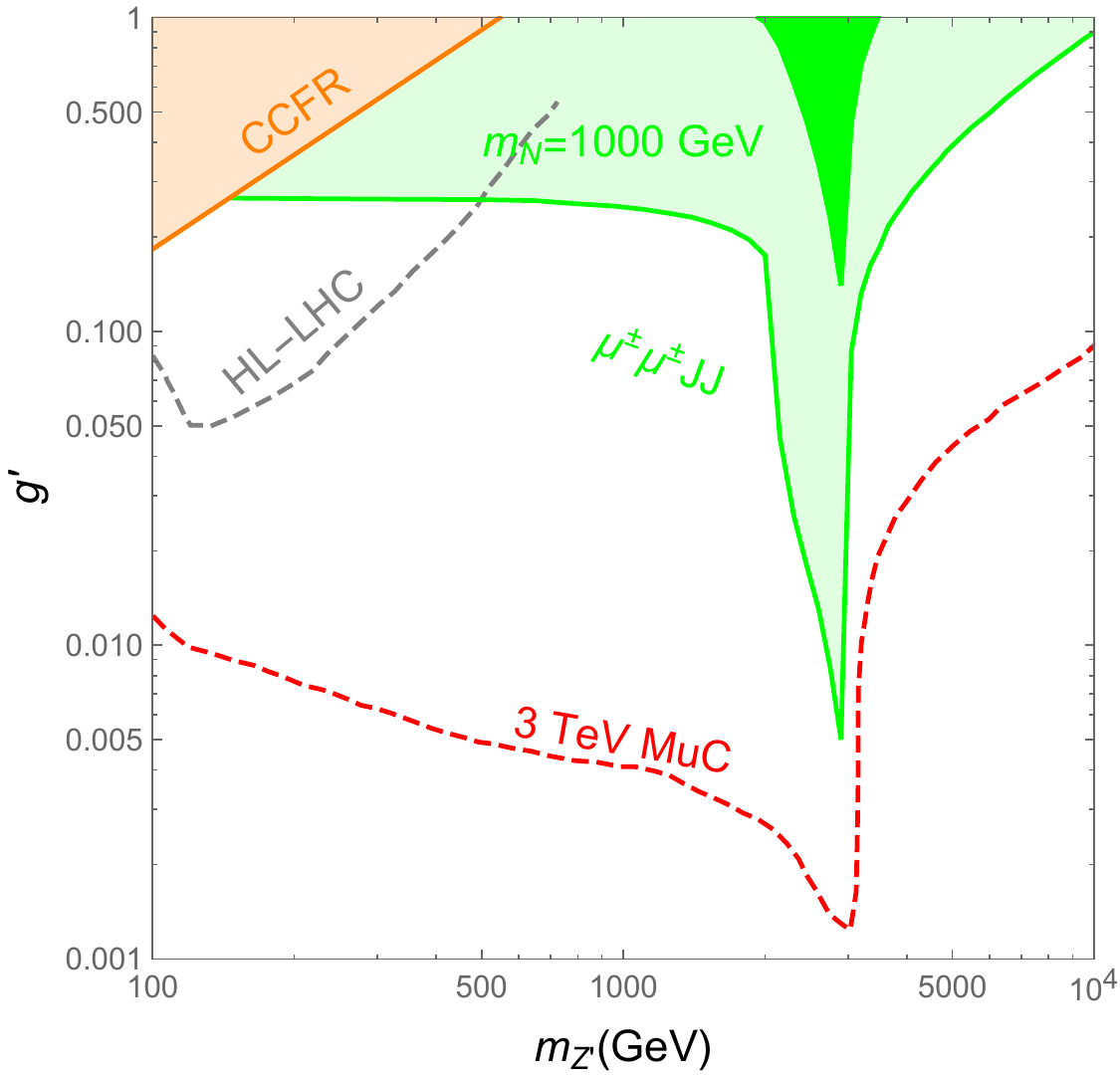}
		\includegraphics[width=0.45\linewidth]{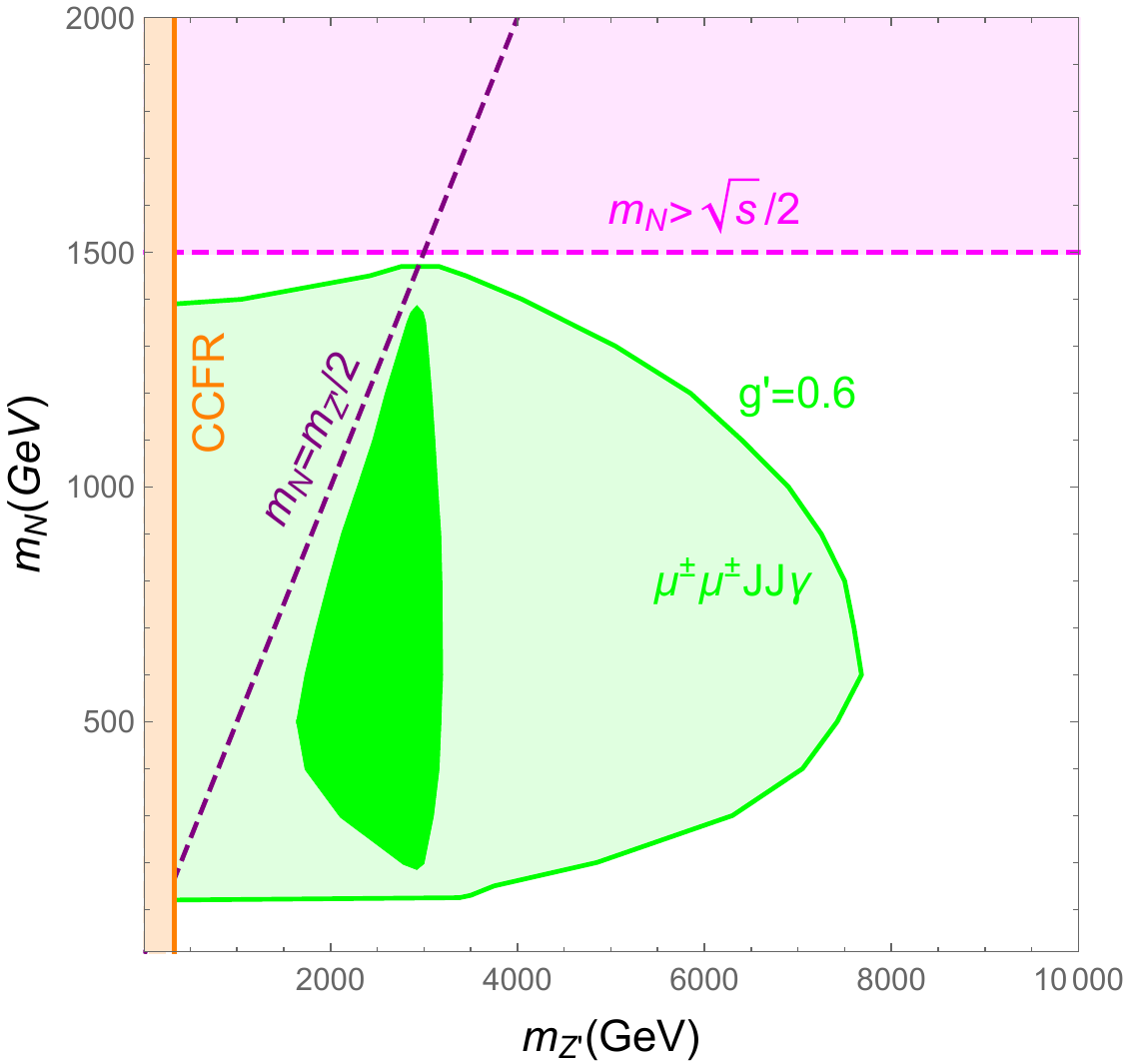}
	\end{center}
	\caption{Same as Figure~\ref{FIG:SC1}, but for the sensitivity of $\mu^+\mu^-JJ\gamma$ signal.}
	\label{FIG:SC2}
\end{figure}

The sensitivity of the $\mu^\pm \mu^\pm JJ\gamma$ signature at the 3 TeV muon collider is explored with the cuts in Table~\ref{Tab:NmA}, and Figure \ref{FIG:SC2} depicts the results. The up-left panel shows the sensitivity region with fixed mass relation $m_N=m_{Z'}/3$, so the decay channel $Z'\to NN$ is always allowed. In this way, we might probe $m_{Z'}$ in the range of [340, 4300]~GeV. Compared with the $\mu^\pm \mu^\pm JJ$ signature, we find that this signal is more promising approximately in the region of [400, 3000] GeV. For example, with $m_{Z'}=2000$ GeV, the $\mu^\pm \mu^\pm JJ\gamma$ signature could probe $g'\gtrsim0.02$, while the $\mu^\pm \mu^\pm JJ$ signature could only reach $g'\sim0.1$. Near $m_{Z'}\sim \sqrt{s}=3$ TeV, we even can reveal $g'\gtrsim0.006$. When $m_{Z'}$ is larger than the collision energy $\sqrt{s}$, the associated photon with the off-shell contribution of $Z'$ makes the $\mu^\pm \mu^\pm JJ\gamma$ signature less detectable than the $\mu^\pm \mu^\pm JJ$ signal.

The up-right panel of Figure \ref{FIG:SC2} shows the sensitivity region with fixed $m_N=1000$ GeV. When $m_{Z'}\leq 2 m_N$, the decay channel $Z'\to NN$ is not allowed, so production of $NN\gamma$ is through the off-shell $Z'$. With nearly a constant cross section, the $\mu^\pm \mu^\pm JJ\gamma$ signature could test $g'\gtrsim0.26$ for $m_{Z'}\leq 2000$ GeV, which is clearly less promising than the $\mu^\pm \mu^\pm JJ$ signature. However, once the decay mode $Z'\to NN$ is open, the cross section of $NN\gamma$ is enhanced, so the $\mu^\pm \mu^\pm JJ\gamma$ signal becomes quite promising. According to our simulation, we find that this signature could probe $g'\gtrsim0.15$ with only 1 fb$^{-1}$ data when 2000 GeV $\lesssim m_{Z'}\lesssim\sqrt{s}$.

The sensitivity region in the $m_N-m_{Z'}$ with fixed $g'=0.6$ is shown in the down panel of Figure \ref{FIG:SC2}. With 1 fb$^{-1}$ data, a large portion of the parameter space with $2m_N\lesssim m_{Z'}\lesssim \sqrt{s}$ could be tested. With a similar reason as the $\mu^\pm \mu^\pm JJ$ signature, the most sensitive $m_N$ of the $\mu^\pm \mu^\pm JJ\gamma$ signature is also approximately 500 GeV. Outside the above resonance $Z'$ region, the significance of the  $\mu^\pm \mu^\pm JJ\gamma$ signature decreases due to the associated photon. Although less promising than the $\mu^\pm \mu^\pm JJ$ signature, the $\mu^\pm \mu^\pm JJ\gamma$ signal could probe the region with $m_{Z'}\lesssim7.6$ TeV and 120 GeV $\lesssim m_N\lesssim1470$ GeV at the 3 TeV muon collider with an integrated luminosity of 1000 fb$^{-1}$. 

\section{Conclusion}\label{SEC:CL}

Heavy neutral leptons $N$ are introduced in the type-I seesaw mechanism to generate tiny neutrino masses. While electroweak scale $N$ is extensively studied at colliders, a relatively large mixing parameter $V_{\ell N}$ is usually required. In this paper, we consider the gauged $U(1)_{L_\mu-L_\tau}$ extension of the type-I seesaw mechanism, where three heavy neutral lepton $N_e,N_\nu,N_\tau$ with $U(1)_{L_\mu-L_\tau}$ charge $(0,1,-1)$ are introduced. Charged under the new gauged  $U(1)_{L_\mu-L_\tau}$ symmetry, there are new production channels of the heavy neutral lepton at colliders. We then investigate the lepton number violation signatures of heavy neutral lepton $N$ in the gauged $U(1)_{L_\mu-L_\tau}$ at the 3~TeV muon collider.

Mediated by the new gauge boson $Z'$, the heavy neutral lepton $N$ can be pair produced at the muon collider via the process $\mu^+\mu^-\to Z^{\prime *}\to NN$ and $\mu^+\mu^-\to Z^{\prime (*)}\gamma\to NN \gamma$. Cross sections of these processes are determined by the new gauge coupling $g'$, gauge boson mass $m_{Z'}$ and heavy neutral lepton mass $m_N$, which are independent of the mixing parameter $V_{\ell N}$. While both processes have a maximum cross section at $m_{Z'}\simeq \sqrt{s}$, the former one is more promising when $m_{Z'}>\sqrt{s}$. For lighter new gauge boson satisfying $2m_N<m_{Z'}<\sqrt{s}$, $Z'$ is produced on-shell with associated photon, and the decay mode $Z'\to NN$ is also allowed. In this way, the cross section of $\mu^+\mu^-\to NN\gamma$ can be enhanced.

Cascade decays of heavy neutral lepton can raise various interesting signatures. Provided the Majorana nature of heavy neutral lepton in the seesaw model, we focus on the lepton number violation signatures in this study. For illustration, we further assume that the heavy neutral lepton $N$ preferentially couples to muon via mixing with light neutrinos. So the dominant decay mode of heavy neutral lepton is $N\to \mu^\pm W^\mp$. The hadronic decays of $W$ boson are considered to reconstruct $m_N$, which are treated as one fat-jet $J$. The explicit signatures are $\mu^+\mu^-\to NN\to \mu^\pm\mu^\pm JJ$ and $\mu^+\mu^-\to NN\gamma\to \mu^\pm\mu^\pm JJ\gamma$.

With relatively clean backgrounds, a cut-based analysis is then performed, which indicates that a large part of the viable parameter space is within the reach of the 3 TeV muon collider. For example, the $\mu^\pm\mu^\pm JJ$ could probe $m_{Z'}$ in the range of [330, 4500]~GeV with fixed mass relation $m_{N}=m_{Z'}/3$ and an integrated luminosity of 1000~fb$^{-1}$. Around the resonance region $m_{Z'}\sim\sqrt{s}$, the gauge coupling $g'$ could be down to about 0.03. We can probe $g'\gtrsim0.15$ with $m_N=1000$~GeV and light $Z'$. On the other hand, this signal could probe $m_{Z'}\lesssim 12.5$~TeV with  $g'=0.6$. Meanwhile, if the gauge boson mass satisfies $2 m_N<m_{Z'}<\sqrt{s}$, the $\mu^\pm\mu^\pm JJ\gamma$ signature would be more promising than the $\mu^\pm\mu^\pm JJ$ signature. It is also notable that around the resonance region $m_{Z'}\sim\sqrt{s}$, both signatures could probe sizable parameter space even with only 1 fb$^{-1}$ data for $g'=0.6$.

\section*{Acknowledgments}
This work is supported by the National Natural Science Foundation of China under Grant  No. 12375074 and 11805081, Natural Science Foundation of Shandong Province under Grant No. ZR2019QA021.

\bibliographystyle{JHEP}


\end{document}